\documentclass[12pt]{article}
\pdfoutput=1

\usepackage{amssymb}
\usepackage{amsmath}
\usepackage{bbold}
\usepackage{color}
\usepackage{colordvi}
\usepackage{fancybox}
\usepackage[footnotesize]{caption2}
\usepackage{graphicx}
\usepackage[center,footnotesize,hang]{subfigure}
\usepackage{bbm}
\usepackage{multirow}
\usepackage{array}
\usepackage{arydshln}
\usepackage[colorlinks]{hyperref}
\usepackage{enumitem}  
\newcommand{\PreserveBackslash}[1]{\let\temp=\\#1\let\\=\temp}
\newcolumntype{C}[1]{>{\PreserveBackslash\centering}p{#1}}
\newcolumntype{R}[1]{>{\PreserveBackslash\raggedleft}p{#1}}
\newcolumntype{L}[1]{>{\PreserveBackslash\raggedright}p{#1}}
\newcommand{\cleqn}{\setcounter{equation}{0}}
\allowdisplaybreaks \allowdisplaybreaks[2]

\begin{document}

\title{
\begin{flushright}
\ \\[-10mm]
\begin{minipage}{0.2\linewidth}
\normalsize
\end{minipage}
\end{flushright}
{\Large \bf
Generalized CP and $\Delta (3n^2)$ Family Symmetry for Semi-Direct Predictions of the PMNS Matrix
\\[2mm]}}

\date{}

\author{
Gui-Jun~Ding$^{1}$\footnote{E-mail: {\tt
dinggj@ustc.edu.cn}},  \
Stephen~F.~King$^{2}$\footnote{E-mail: {\tt king@soton.ac.uk}} \
\\*[20pt]
\centerline{
\begin{minipage}{\linewidth}
\begin{center}
$^1${\it \small
Department of Modern Physics, University of Science and Technology of
China,\\
Hefei, Anhui 230026, China}\\[2mm]
$^2${\it \small
Physics and Astronomy,
University of Southampton,
Southampton, SO17 1BJ, U.K.}\\
\end{center}
\end{minipage}}
\\[10mm]}
\maketitle
\thispagestyle{empty}

\begin{abstract}
\noindent
The generalized CP transformations can only be consistently defined in the context of $\Delta(3n^2)$ lepton symmetry if a certain subset of irreducible representations are present in a model. We perform a comprehensive analysis of the possible automorphisms and the corresponding CP transformations of the $\Delta(3n^2)$ group. It is sufficient to only consider three automorphisms if $n$ is not divisible by 3 while additional eight types of CP transformations could be imposed for the case of $n$ divisible by 3. We study the lepton mixing patterns which can be derived from the $\Delta(3n^2)$ family symmetry and generalized CP in the semi-direct approach. The PMNS matrix is determined to be the trimaximal pattern for all the possible CP transformations, and it can only take two distinct forms.

\end{abstract}

\newpage

\section{\label{sec:introduction}Introduction}
\indent

A large number of neutrino oscillation experiments have established that neutrinos are massive and the flavor mixing between the different lepton generations are described by the Pontecorvo-Maki-Nakagawa-Sakata (PMNS) matrix which are parameterized by three mixing angles $\theta_{12}$, $\theta_{13}$ and $\theta_{23}$, one Dirac CP violating phase $\delta_{CP}$ and two additional Majorana phases $\alpha_{21}$ and $\alpha_{31}$ if neutrinos are Majorana particles~\cite{Agashe:2014kda}. So far all the three mixing angles have been measured~\cite{Gonzalez-Garcia:2014bfa,Capozzi:2013csa,Forero:2014bxa}. However, all the CP phases are unconstrained at $3\sigma$ level, although weak evidence for a maximal Dirac phase $\delta_{CP}\simeq3\pi/2$ have been observed by the T2K collaboration~\cite{Abe:2015awa}.

The origin of neutrino masses and mixing parameters is a longstanding open question in particle physics. The discrete symmetry approach has been widely explored to understanding the observed pattern of neutrino mixing, see Refs.~\cite{Altarelli:2010gt,Ishimori:2010au,King:2013eh,King:2014nza,King:2015aea}
for review. In this approach, it is assumed that there is a finite, non-Abelian flavor symmetry $G_f$ at some high energy scale, and the standard model particles are assigned to its irreducible representations. Subsequently $G_{f}$ is broken down to different residual subgroups $G_{\nu}$ and $G_{l}$ in the neutrino and charged lepton sectors respectively. The PMNS matrix is completely fixed by the mismatch of the embedding of the residual subgroups $G_{\nu}$ and $G_{l}$ into the flavor symmetry group $G_{f}$. It is not necessary to specify the breaking mechanism in order to predict the lepton mixing parameters.

In general, there are three possible implementations of flavor symmetries, known as ``direct'', ``semi-direct'' and ``indirect''~\cite{King:2013eh}. In the ``direct'' approach, all the low energy residual symmetry of the neutrino mass matrix is a subgroup of $G_{f}$ such that both mixing angles and Dirac phase would be predicted to be some constant values. For neutrinos being either Majorana particles or Dirac particles, it is found that only possible mixing pattern which would agree with current data is the trimaximal mixing, $\delta_{CP}$ is conserved, and the flavor symmetry group $G_{f}$ is larger~\cite{Holthausen:2012wt,Fonseca:2014koa,Yao:2015dwa} .
In the ``semi-direct'' approach, the symmetry of the neutrino mass matrix, typically $Z_2$, is partially contained in $G_{f}$. As a consequence, only one column of the PMNS matrix could be fixed and an arbitrary unitary rotation in the degenerate subspace of the neutrino residual symmetry is allowed. In this way, the experimental data in particular non-vanishing $\theta_{13}$ can be easily accommodated without needing larger flavor symmetry group $G_{f}$. Recently this approach is extended to include the generalized CP as a symmetry~\cite{Feruglio:2012cw,Holthausen:2012dk}. This can lead to a very predictive scenario in which all the mixing angles and CP phases only depend on one single real parameter~\cite{Feruglio:2012cw}. The generalized CP symmetry was firstly explored in the context of continuous gauge groups~\cite{Ecker:1981wv,Grimus:1995zi}, and the so-called $\mu\tau$ reflection is a typical predictive generalized CP transformations~\cite{Harrison:2002kp,Grimus:2003yn,Farzan:2006vj}.
In order to give a consistent definition of generalized CP transformations in the presence of discrete flavor symmetry, certain consistency condition must be fulfilled~\cite{Holthausen:2012dk,Chen:2014tpa}. The relationship between neutrino mixing and CP symmetry has been clarified~\cite{Chen:2014wxa,Chen:2015nha,Everett:2015oka}, and the master formula to reconstruct the PMNS matrix from any given remnant CP transformation has been derived~\cite{Chen:2014wxa,Chen:2015nha}. The phenomenological predictions and model building of combining discrete flavor symmetry with generalized CP have already been studied for a number of groups in the literature including $A_4$~\cite{Ding:2013bpa}, $S_4$~\cite{Feruglio:2012cw,Ding:2013hpa,Feruglio:2013hia,Luhn:2013lkn,Li:2013jya,Li:2014eia}, $A_5$~\cite{Li:2015jxa,DiIura:2015kfa,Ballett:2015wia,Turner:2015uta},
$\Delta(27)$~\cite{Branco:1983tn,Branco:2015gna},
$\Delta(48)$~\cite{Ding:2013nsa,Ding:2014hva}, $\Delta(96)$~\cite{Ding:2014ssa} and the infnite series of finite
groups $\Delta(3n^2)$~\cite{Hagedorn:2014wha} and $\Delta(6n^2)$~\cite{Hagedorn:2014wha,King:2014rwa,Ding:2014ora}.

Based on our previous discussions of smaller $\Delta(3n^2)$ group $A_4\cong\Delta(3\cdot2^2)$~\cite{Ding:2013bpa} and $\Delta(48)\equiv\Delta(3\cdot4^2)$~\cite{Ding:2013nsa,Ding:2014hva} combined with generalized CP, we shall perform a comprehensive study of all possible automorphisms of the $\Delta(3n^2)$ group for general $n$ and corresponding CP symmetry and its predictions for the lepton flavor mixing in the ``semi-direct'' framework. The distinction between the present work and the $\Delta(3n^2)$ part of Ref.~\cite{Hagedorn:2014wha} is that the latter focuses on a particular set of CP transformations represented by $\mu\tau$ reflection and it only considers the $\Delta(3n^2)$ groups with $n$ not divisible by three, whereas our dedicated analysis here is more general.

This paper is structured as follows. In section~\ref{sec:group_theory}, the basic aspects of the $\Delta(3n^2)$ group are reviewed. A complete classification of the permissible automorphisms of $\Delta(3n^2)$ group and explicit form of the corresponding generalized CP transformations are presented in section~\ref{sec:GCP_def}. We study the possible mixing patterns which can be achieved in the ``semi-direct'' approach from the $\Delta(3n^2)$ family symmetry and generalized CP in section~\ref{sec:mixing_semi-direct}. Finally we summarize and present our conclusions in section~\ref{sec:conclusion}. The Clebsch-Gordan coefficients of the $\Delta(3n^2)$ group are collected in Appendix~\ref{sec:appendix_A}, they would be useful in concrete model building.

\section{\label{sec:group_theory}The group theory of $\Delta(3n^2)$}
\indent

$\Delta(3n^2)$ is non-abelian finite subgroup of $SU(3)$ of order $3n^2$. $\Delta(3n^2)$ is isomorphic to $(Z_n\times Z_n)\rtimes Z_3$, $Z_n\times Z_n$ forms a normal, abelian subgroup of $\Delta(3n^2)$, generated by the elements $c$ and $d$, and the $Z_3$ generator is written by $a$. These three generators $a$, $b$ and $c$ fulfill the following rules~\cite{Luhn:2007uq}
\begin{eqnarray}
\nonumber& a^3=1,\quad c^n=d^n=1,\quad cd=dc,\\
\label{eq:Delta3N2_generators}&aca^{-1}=c^{-1}d^{-1},\qquad ada^{-1}=c\,.
\end{eqnarray}
Since we have $d=a^{-1}ca$, only two generators $a$ and $c$ are independent. All the group elements can be written as
\begin{equation}
g=a^{\alpha}c^{\gamma}d^{\delta}\,,
\end{equation}
where $\alpha=0,1,2$ and $\gamma, \delta=0,1,2,\ldots n-1$. The $\Delta(3n^2)$ group have been thoroughly studied in Ref.~\cite{Luhn:2007uq}. In the following, we shall review the basic aspects which are relevant to our present work. Depending whether or not $n$ is divisible by 3, the $3n^2$ elements of $\Delta(3n^2)$ can be categorized into the following conjugacy classes:
\begin{itemize}
\item{$n\neq3\, \mathbb{Z}$}
\begin{eqnarray}
\nonumber 1C_1&=&\left\{1\right\},\\
\nonumber 3C^{(\rho, \sigma)}_1&=&\left\{c^{\rho}d^{\sigma}, c^{-\rho+\sigma}d^{-\rho}, c^{-\sigma}d^{\rho-\sigma}\right\},\\
\nonumber n^2C_2&=&\left\{ac^{x}d^{y}|c,y=0, 1, \ldots, n-1\right\}\,,\\
 n^2C_3&=&\left\{a^2c^{x}d^{y}|c,y=0, 1, \ldots, n-1\right\}\,,
\end{eqnarray}
where $\rho, \sigma=0, 1, \ldots, n-1$, and the prefix before each class represents the number of elements in the class. Although $\rho$ and $\sigma$ can take $n^2-1$ possible values, the triples $(\rho, \sigma)$, $(-\rho+\sigma, -\rho)$ and $(-\sigma, \rho-\sigma)$ lead to the same class, we have $(n^2-1)/3$ classes of the form $3C^{(\rho, \sigma)}_1$. Notice that all the elements in both $n^2C_2$ and $n^2C_3$ are of order three.

\item{$n=3\, \mathbb{Z}$}

\begin{eqnarray}
\nonumber 1C_1&=&\left\{1\right\},\\
\nonumber 1C^{(\rho)}_{1}&=&\left\{c^{\rho}d^{-\rho}\right\},\quad \rho=\frac{n}{3},~\frac{2n}{3}\,,\\
\nonumber 3C^{(\rho, \sigma)}_1&=&\left\{c^{\rho}d^{\sigma}, c^{-\rho+\sigma}d^{-\rho}, c^{-\sigma}d^{\rho-\sigma}\right\},\quad (\rho, \sigma)\neq\big(\frac{n}{3}, \frac{2n}{3}\big),~\big(\frac{2n}{3}, \frac{n}{3}\big)\,,\\
\nonumber \frac{n^2}{3}C^{(\tau)}_{2}&=&\Big\{ac^{\tau-y-3x}d^{y}|x=0, 1, \ldots, \frac{n-3}{3}; y=0, 1, \ldots, n-1\Big\}\,,\\
\frac{n^2}{3}C^{(\tau)}_3&=&\Big\{a^2c^{\tau-y-3x}d^{y}|x=0, 1, \ldots, \frac{n-3}{3}; y=0, 1, \ldots, n-1\Big\}\,,
\end{eqnarray}
where $\tau=0, 1, 2$, and we have $(n^2-3)/3$ three-element classes of form $3C^{(\rho, \sigma)}_1$ in this case. Note that the $Z_3$ subgroup $\left\{1, c^{\frac{n}{3}}d^{\frac{2n}{3}}, c^{\frac{2n}{3}}d^{\frac{n}{3}}\right\}$ is the center of the group. The elements in the conjugacy classes $\frac{n^2}{3}C^{(\tau)}_{2}$ and $\frac{n^2}{3}C^{(\tau)}_3$ are of order three as well.

\end{itemize}

The irreducible representation of the $\Delta(3n^2)$ group is either one-dimensional or three-dimensional. If $n$ isn't divided by three, $\Delta(3n^2)$ has three singlet representations and $(n^2-1)/3$ triplet representations while it has nine singlet representations and $(n^2-3)/3$ triplet representations if $n$ can be divided by three. The explicit form of the representation matrices are as follows.
\begin{itemize}
\item{One-dimensional representations}
\begin{eqnarray}
\nonumber&&\mathbf{1}_{r}:~a=\omega^{r},\quad c=d=1,~~\text{for}~~n\neq3\, \mathbb{Z}\,,\\
&&\mathbf{1}_{r, s}:~a=\omega^{r},\quad c=d=\omega^{s},~~\text{for}~~n=3\, \mathbb{Z}\,,
\end{eqnarray}
with $r, s=0, 1, 2$ and $\omega\equiv e^{2\pi i/3}$.

\item{Three-dimensional representations}

\begin{equation}
\label{eq:3-dim_rep}\mathbf{3}_{\widetilde{(k, l)}}:~a=\left(\begin{array}{ccc}
0   &   1   &  0 \\
0   &   0   &  1 \\
1   &   0   &  0
\end{array}
\right),\quad c=\left(\begin{array}{ccc}
\eta^{l}  &   0    &   0 \\
0         &   \eta^{k}  &   0 \\
0   &   0   &  \eta^{-k-l}
\end{array}
\right),\quad  d=\left(\begin{array}{ccc}
\eta^{-k-l}   &   0   &   0  \\
0         &   \eta^{l}   &  0  \\
0     &    0      &   \eta^{k}
\end{array}\right)\,,
\end{equation}
where $\eta\equiv e^{2\pi i/n}$ and $k, l=0, 1, \ldots, n-1$ with $(k, l)\neq(0,0)$. The notation $\widetilde{\phantom{wa}}$ denotes the mapping
\begin{equation}
\widetilde{\left(\begin{array}{c} k \\ l \end{array}\right)} ~\longmapsto
\left(\begin{array}{c} k \\ l \end{array}\right), ~~\text{or}~~
\left(\begin{array}{c} -k-l \\ k \end{array}\right), ~~\text{or}~~
\left(\begin{array}{c} l \\ -k-l \end{array}\right)\,.
\end{equation}
The reason is that the representations labeled by the three pairs $(k, l)$, $(-k-l, k)$ and $(l, -k-l)$ are equivalent. Notice that the representation matrix of the generator $a$ is the same in all triplet representations. $\mathbf{3}_{(k, l)}$ and $\mathbf{3}_{(k^{\prime}, l^{\prime})}$ are complex conjugate of each other if $k+k^{\prime}=n$ and $l+l^{\prime}=n$. One can check that $\mathbf{3}_{(1, n-1)}$ is always a faithful representation of $\Delta(3n^2)$ no matter whether $n$ is divisible by three or not. On the other hand, $\mathbf{3}_{(1, 1)}$ is a faithful representation for $n\neq3\, \mathbb{Z}$ while it is not for $n=3\, \mathbb{Z}$ as we have $\rho_{\mathbf{3}_{(1,1)}}(c^{\frac{n}{3}}d^{\frac{2n}{3}})=\rho_{\mathbf{3}_{(1,1)}}(c^{\frac{2n}{3}}d^{\frac{n}{3}})=1$.

\end{itemize}

The $\Delta(6n^2)$ group can be obtained from $\Delta(3n^2)$ by including another generator $b$ which fulfills the following relations~\cite{Escobar:2008vc}
\begin{equation}
b^2=(ab)^2=1,\quad bcb^{-1}=d^{-1},\quad bdb^{-1}=c^{-1}\,.
\end{equation}
The element of $\Delta(6n^2)$ can be expressed as $g=a^{\alpha}b^{\beta}c^{\gamma}d^{\delta}$, where $\alpha=0, 1,2 $, $\beta=0, 1$ and $\gamma, \delta=0, 1, \ldots, n-1$. $\Delta(6n^2)$ group has $2n-2$ three-dimensional irreducible representations denoted as $\mathbf{3}_{1, k}$ and $\mathbf{3}_{2, k}$, and the representation matrices of the generators are given by~\cite{Escobar:2008vc}
\begin{equation}
\label{eq:rep_Delta(6n2)}a=\begin{pmatrix}
0  &~  1  &~  0  \\
0  &~  0  &~  1  \\
1  &~  0  &~  0
\end{pmatrix},\quad
b=\pm\begin{pmatrix}
0   &~   0    &~  1 \\
0   &~   1    &~  0 \\
1   &~   0    &~  0
\end{pmatrix},\quad
c=\begin{pmatrix}
\eta^{-k}  &~   0  &~  0 \\
0  &~   \eta^{k}  &~  0  \\
0  &~   0    &~   1
\end{pmatrix},\quad
d=\begin{pmatrix}
1   &~   0   &~  0  \\
0   &~  \eta^{-k}   &~  0  \\
0   &~   0      &~  \eta^{k}
\end{pmatrix}\,,
\end{equation}
with $k=1, 2,\ldots, n-1$. We see that $\mathbf{3}_{1, k}$ differs from $\mathbf{3}_{2, k}$ in the overall sign of the generator $b$ and the representation matrices for the generators $a$, $c$ and $d$ coincide with those of $\Delta(3n^2)$ in the triplet representation $\mathbf{3}_{k, n-k}$.


\section{\label{sec:GCP_def}Automorphism of $\Delta(3n^2)$ group and the generalized CP transformations}

The interplay between flavor symmetry and generalized CP symmetry has been comprehensively discussed~\cite{Ecker:1981wv,Grimus:1995zi,Holthausen:2012dk,Feruglio:2012cw,Ding:2013hpa,Chen:2014tpa}. It is nontrivial to consistently define a generalized CP symmetry in the presence of a flavor symmetry $G_{f}$. The generalized CP symmetry has to be compatible with the family symmetry $G_{f}$, and the following consistency condition should be satisfied~\cite{Ecker:1981wv,Grimus:1995zi,Holthausen:2012dk,Feruglio:2012cw,Ding:2013hpa}:
\begin{equation}
\label{eq:consistency_condition}X_{\mathbf{r}}\rho^{*}_{\mathbf{r}}(g)X^{\dagger}_{\mathbf{r}}=\rho_{\mathbf{r}}(g^{\prime}),\quad g,g^{\prime}\in G_{f}\,,
\end{equation}
where $\rho_{\mathbf{r}}(g)$ denotes the representation matrix of the element $g$ in the irreducible representation $\mathbf{r}$, and $X_{\mathbf{r}}$ is the so-called generalized CP transformation. Hence the generalized CP transformation is related to an automorphism $\mathfrak{u}$ which maps $g$ into $g^{\prime}$. Furthermore, it was shown that the physical CP transformations have to be given by class-inverting automorphism of $G_f$~\cite{Chen:2014tpa}. In this work, we shall investigate the $\Delta(3n^2)$ series as the family symmetry group. With the help of the computer algebra program system \texttt{GAP}~\cite{GAP4:2011}, we have studied the automorphism group of $\Delta(3n^2)$ with $n=1, 2, \ldots, 26$. Note that the $\Delta(3n^2)$ groups with $n>26$ are not available in \texttt{GAP} at present. We find that the $\Delta(3n^2)$ group generally doesn't have class-inverting automorphism except the first two smallest ones $\Delta(3\cdot1^2)\cong Z_3$ and $\Delta(3\cdot2^2)\cong A_4$. The same conclusion has been reached in Ref.~\cite{Hagedorn:2014wha}. Therefore generically one is not able to introduce physically well-defined CP transformations within the context of $\Delta(3n^2)$ flavor symmetry. However, if a concrete model contains only a subset of irreducible representations for which an automorphism $\mathfrak{u}$ exists and it exchanges each of these representations with the complex conjugate, one can impose the generalized CP transformation corresponding to $\mathfrak{u}$ to be a symmetry. In the following, we shall perform a comprehensive study of all possible admissible CP transformations. Firstly we shall concentrate on the case of $n\neq3\, \mathbb{Z}$.

Obviously the elements $g$ and $g^{\prime}$ in the consistency equation of Eq.~\eqref{eq:consistency_condition} should be of the same order, and the characters of $g$ and $g^{\prime}$ are conjugate. As a result, the generator $a$ could be mapped into an element belonging to the conjugacy class $n^2C_2$ or $n^2C_3$. Without loss of generality, it is sufficient to consider two scenarios that $a$ is mapped to $a$ or $a^2$, since the automorphism which maps $a$ into other element of $n^2C_2$ and $n^2C_3$ can be generated by performing group conjugation further. In a similar fashion, another generator $c$ should be mapped to an element in the conjugacy class $3C^{(\rho, \sigma)}_1$. Notice that the representation matrix of the generator $a$ is of the same form for all the triplet representations and representation matrices of the class $3C^{(\rho, \sigma)}_1$ are diagonal in our working basis. By studying the consistency equation of Eq.~\eqref{eq:consistency_condition} for both generators $a$ and $c$, we find that the CP transformation for the automorphisms mapping $a$ to $a^2$ is always of the form
\begin{equation}
\label{eq:a_to_a2}X_{0}=\begin{pmatrix}
0   ~&~   1   ~&~   0 \\
1   ~&~   0   ~&~   0 \\
0   ~&~   0   ~&~   1
\end{pmatrix}
\end{equation}
up to flavor symmetry transformations of $a$ and $a^2$. On the other hand, the CP transformation for the automorphisms mapping $a$ to $a$ is always an identity matrix up to flavor transformation of $a$ and $a^2$, i.e.
\begin{equation}
\label{eq:a_to_a2}X_{0}=\begin{pmatrix}
1   ~&~   0   ~&~   0 \\
0   ~&~   1   ~&~   0 \\
0   ~&~   0   ~&~   1
\end{pmatrix}\,.
\end{equation}
As a consequence, all possible admissible CP transformations of $\Delta(3n^2)$ can be fixed by considering three representative automorphisms. The first one is \begin{equation}
a\stackrel{\mathfrak{u_1}}{\longmapsto}a^{-1},\quad c\stackrel{\mathfrak{u_1}}{\longmapsto}c^{-1},\quad d\stackrel{\mathfrak{u_1}}{\longmapsto}cd\,.
\end{equation}
Under the action of this automorphism, the different conjugacy classes are transformed into
\begin{equation}
3C^{(\rho, \sigma)}_1\stackrel{\mathfrak{u_1}}{\longmapsto}3C^{(-\sigma, -\rho)}_1,\qquad n^2C_2\stackrel{\mathfrak{u_1}}{\longleftrightarrow}n^2C_3\,.
\end{equation}
Notice that $3C^{(-\sigma, -\rho)}_1$ is generically not the inverse class of $3C^{(\rho, \sigma)}_1$ unless $\rho=\sigma$, $\rho=0$ or $\sigma=0$. Therefore $\mathfrak{u}_1$ is not a class-inverting automorphism. This automorphism $\mathfrak{u}_1$ acts on the irreducible representations of $\Delta(3n^2)$ as follows
\begin{equation}
\mathbf{1}_{r}\stackrel{\mathfrak{u_1}}{\longmapsto}\mathbf{1}_{-r},\qquad \mathbf{3}_{\widetilde{(k, l)}}\stackrel{\mathfrak{u_1}}{\longmapsto}\mathbf{3}_{\widetilde{(-l,-k)}}\,,
\end{equation}
where $\mathbf{1}_{r}\stackrel{\mathfrak{u_1}}{\longmapsto}\mathbf{1}_{-r}$ is to be read as $\rho_{\mathbf{1}_{-r}}=\rho_{\mathbf{1}_{r}}\circ u_1$ etc. Therefore the three singlets $\mathbf{1}_{r}$ with $r=0, 1, 2$ and the triplet representations $\mathbf{3}_{\widetilde{(k, k)}}$ with $k=1, 2, \ldots, n-1$ can be present while other three-dimensional representations should be absent in order to consistently define the generalized CP transformations corresponding to the automorphism $\mathfrak{u}_1$. In the case of $n=2$, the flavor symmetry group $\Delta(3\cdot2^2)\cong A_4$ has four irreducible representations $\mathbf{1}_0$, $\mathbf{1}_1$, $\mathbf{1}_2$ and $\mathbf{3}_{(1,1)}\cong\mathbf{3}_{(0,1)}\cong\mathbf{3}_{(1,0)}$, the CP transformation associated with the $\mathfrak{u}_1$ can be imposed onto an $A_4$ model regardless of the field content. The explicit form of the CP transformation related to $\mathfrak{u}_1$ is determined by the following consistency equations:
\begin{eqnarray}
\nonumber&&X_{\mathfrak{u}_1}\rho^{\ast}_{\mathbf{3}_{(k, k)}}(a)X^{\dagger}_{\mathfrak{u}_1}=\rho_{\mathbf{3}_{(k, k)}}(\mathfrak{u}_1(a))=\rho_{\mathbf{3}_{(k, k)}}(a^2),\\
&&X_{\mathfrak{u}_1}\rho^{\ast}_{\mathbf{3}_{(k, k)}}(c)X^{\dagger}_{\mathfrak{u}_1}=\rho_{\mathbf{3}_{(k, k)}}(\mathfrak{u}_1(c))=\rho_{\mathbf{3}_{(k, k)}}(c^{-1})\,,
\end{eqnarray}
which yields
\begin{equation}
X_{\mathfrak{u}_1}=\begin{pmatrix}
0  &~  1  &~  0  \\
1  &~  0  &~  0  \\
0  &~  0  &~  1
\end{pmatrix}\,.
\end{equation}
Furthermore, including the inner automorphism $\text{conj}(g): h\rightarrow ghg^{-1}$ with $g, h\in\Delta(3n^2)$, which is described by a group transformation~\cite{Holthausen:2012dk,Feruglio:2012cw,Ding:2013hpa}, we find the following CP transformations are also admissible
\begin{equation}
\rho_{\mathbf{3}_{(k, k)}}(a^{\alpha}c^{\gamma}d^{\delta})X_{\mathfrak{u}_1}
\end{equation}
where $\alpha=0,1,2$ and $\gamma, \delta=0,1,2,\ldots n-1$. The associated automorphism $\text{conj}(a^{\alpha}c^{\gamma}d^{\delta})\mathfrak{u}_1$ can map the generators $a$ and $c$ into any element of $n^2C_3$ and $3C^{(-1, 0)}_1$ respectively. The second representative automorphism $\mathfrak{u_2}$ is  generated by
\begin{equation}
a\stackrel{\mathfrak{u_2}}{\longmapsto}a^{-1},\quad c\stackrel{\mathfrak{u_2}}{\longmapsto}c,\quad d\stackrel{\mathfrak{u_2}}{\longmapsto}c^{-1}d^{-1}\,.
\end{equation}
The conjugacy classes are mapped into
\begin{equation}
3C^{(\rho, \sigma)}_1\stackrel{\mathfrak{u_2}}{\longmapsto}3C^{(\sigma, \rho)}_1\;,\quad n^2C_2\stackrel{\mathfrak{u_2}}{\longleftrightarrow}n^2C_3\,,
\end{equation}
This automorphism $\mathfrak{u}_2$ acts on the irreducible representations as
\begin{equation}
\mathbf{1}_{r}\stackrel{\mathfrak{u_2}}{\longmapsto}\mathbf{1}_{-r}\;,\qquad \mathbf{3}_{\widetilde{(k, l)}}\stackrel{\mathfrak{u_2}}{\longmapsto}\mathbf{3}_{\widetilde{(l,k)}}\,,
\end{equation}
Therefore one can impose the CP transformation corresponding to the automorphism $\mathfrak{u}_2$ in the presence of $\Delta(3n^2)$ flavor symmetry if a model contain the irreducible representations $\mathbf{1}_{r}$ ($r=0, 1, 2$) and $\mathbf{3}_{(k, n-k)}$ with $k\neq0$ for which $\mathfrak{u}_2$ exchanges each of these representations by its conjugate.
The CP transformation matrix is a representation of the automorphism $\mathfrak{u}_2$ in the sense of consistency equations:
\begin{eqnarray}
\nonumber&&X_{\mathfrak{u}_2}\rho^{\ast}_{\mathbf{3}_{(k, n-k)}}(a)X^{\dagger}_{\mathfrak{u}_2}=\rho_{\mathbf{3}_{(k, n-k)}}(\mathfrak{u}_2(a))=\rho_{\mathbf{3}_{(k, n-k)}}(a^2),\\
&&X_{\mathfrak{u}_2}\rho^{\ast}_{\mathbf{3}_{(k, n-k)}}(c)X^{\dagger}_{\mathfrak{u}_2}=\rho_{\mathbf{3}_{(k, n-k)}}(\mathfrak{u}_2(c))=\rho_{\mathbf{3}_{(k, n-k)}}(c)\,,
\end{eqnarray}
which lead to
\begin{equation}
X_{\mathfrak{u}_2}=\begin{pmatrix}
0  &~  1  &~  0  \\
1  &~  0  &~  0  \\
0  &~  0  &~  1
\end{pmatrix}\,,
\end{equation}
which is of the same form as $X_{\mathfrak{u}_1}$. By performing a CP transformation followed by a group transformation $\rho_{\mathbf{3}_{(k, n-k)}}(a^{\alpha}c^{\gamma}d^{\delta})$, the matrix
\begin{equation}
\rho_{\mathbf{3}_{(k, n-k)}}(a^{\alpha}c^{\gamma}d^{\delta})X_{\mathfrak{u}_2}
\end{equation}
is also a viable CP transformation, and it can map the generators $a$ and $c$ into any element of $n^2C_3$ and $3C^{(1, 0)}_1$ respectively. Finally the last representative automorphism $\mathfrak{u_3}$ is given by
\begin{equation}
a\stackrel{\mathfrak{u_3}}{\longmapsto}a,\quad c\stackrel{\mathfrak{u_3}}{\longmapsto}c^{-1},\quad d\stackrel{\mathfrak{u_3}}{\longmapsto}d^{-1}\,.
\end{equation}
The action of the automorphism $\mathfrak{u}_3$ on the conjugacy classes of $\Delta(3n^2)$ is as follows
\begin{equation}
3C^{(\rho, \sigma)}_1\stackrel{\mathfrak{u_3}}{\longmapsto}3C^{(-\rho, -\sigma)}_1\;,~ n^2C_2\stackrel{\mathfrak{u_3}}{\longmapsto}n^2C_2\;,~ n^2C_3\stackrel{\mathfrak{u_3}}{\longmapsto}n^2C_3\,.
\end{equation}
This automorphism $\mathfrak{u}_3$ interchanges each three-dimensional representation with its complex conjugate, i.e.
\begin{eqnarray}
\mathbf{1}_{r}\stackrel{\mathfrak{u_3}}{\longmapsto}\mathbf{1}_{r}\;,\qquad \mathbf{3}_{\widetilde{(k, l)}}\stackrel{\mathfrak{u_3}}{\longmapsto}\mathbf{3}_{\widetilde{(-k,-l)}}\,.
\end{eqnarray}
Therefore only the nontrivial singlets $\mathbf{1}_1$ and $\mathbf{1}_2$ should be absent in a model in order to consistently combine the CP transformation corresponding to $\mathfrak{u}_3$ and the $\Delta(3n^2)$ flavor symmetry. Obviously the CP transformation $X_{\mathfrak{u}_3}$ is a unit matrix for all the triplet representations, i.e.
\begin{equation}
X_{\mathfrak{u}_3}=\begin{pmatrix}
1  ~&~   0   ~&~    0  \\
0  ~&~   1   ~&~    0  \\
0  ~&~   0   ~&~    1
\end{pmatrix}\,.
\end{equation}
After considering the inner automorphism, the generalized CP transformations are found to be of the same form as the flavor symmetry transformation $\rho_{\mathbf{3}_{(k, l)}}(a^{\alpha}c^{\gamma}d^{\delta})$ in our working basis, and it can map the generators $a$ and $c$ into any element of $n^2C_2$ and $3C^{(-1, 0)}_1$ respectively. This form of CP transformation has been considered in Ref.~\cite{Hagedorn:2014wha}.

Following the same method, in the case of $n=3\, \mathbb{Z}$, the automorphism of $\Delta(3n^2)$ group can map the generator $a$ into an element of the class $\frac{n^2}{3}C^{(\tau)}_{2}$ or $\frac{n^2}{3}C^{(\tau)}_3$. The generator $c$ should be mapped to an element in the conjugacy class $3C^{(\rho, \sigma)}_1$ except the case of $n=3$ (the flavor symmetry is then $\Delta(27)$). The possible outer automorphisms and the corresponding generalized CP transformations are summarized in Table~\ref{tab:GCP_n=3N}.

We would like to remind that $\Delta(6n^2)$ group has a unique class-inverting outer automorphism for $n\neq3\, \mathbb{Z}$, and the corresponding CP transformations are of the same form as the family symmetry
transformations in the working basis given by Eq.~\eqref{eq:rep_Delta(6n2)}, i.e. $X_{\mathbf{r}}=\rho_{\mathbf{r}}(a^{\alpha}b^{\beta}c^{\gamma}d^{\delta})$~\cite{Ding:2014ora}. In the case of $n=3\, \mathbb{Z}$, the same set of generalized CP transformation can be imposed onto a model if the fields transforming as doublet representations $\mathbf{2}_2$, $\mathbf{2}_3$ and $\mathbf{2}_4$ are not included~\cite{Ding:2014ora}.

\begin{table}[hptb]
\addtolength{\tabcolsep}{-2pt}
\begin{center}
\begin{tabular}{|c|c|c|c|}
\hline\hline
\texttt{Automorphisms}   &   \texttt{Action on rep}   &  \texttt{Rep contents}   &  \texttt{$X_{\mathfrak{u}}$}  \\\hline
 & & &  \\[-0.14in]
$\left(a, c\right)\stackrel{\mathfrak{u_1}}{\longmapsto}\left(a^{-1}, c^{-1}\right)$  & $\begin{array}{c}
\mathbf{1}_{r, s}\stackrel{\mathfrak{u_1}}{\longmapsto}\mathbf{1}_{-r, -s}\\
\mathbf{3}_{\widetilde{(k, l)}}\stackrel{\mathfrak{u_1}}{\longmapsto}\mathbf{3}_{\widetilde{(-l,-k)}}
\end{array}$   &  $\mathbf{1}_{r, s}\,, \mathbf{3}_{(k, k)}$   &  $\begin{pmatrix}
0  &~  1  &~  0  \\
1  &~  0  &~  0  \\
0  &~  0  &~  1
\end{pmatrix}$  \\
 & & &  \\[-0.14in]\hline
 & & &  \\[-0.14in]
$\left(a, c\right)\stackrel{\mathfrak{u_2}}{\longmapsto}\left(a^{-1}, c\right)$  & $\begin{array}{c}
\mathbf{1}_{r, s}\stackrel{\mathfrak{u_2}}{\longmapsto}\mathbf{1}_{-r, s}\\
\mathbf{3}_{\widetilde{(k, l)}}\stackrel{\mathfrak{u_2}}{\longmapsto}\mathbf{3}_{\widetilde{(l,k)}}
\end{array}$   &  $\mathbf{1}_{r, 0}\,, \mathbf{3}_{(k, n-k)}$   &  $\begin{pmatrix}
0  &~  1  &~  0  \\
1  &~  0  &~  0  \\
0  &~  0  &~  1
\end{pmatrix}$  \\
 & & &  \\[-0.14in]\hline
 & & &  \\[-0.14in]
$\left(a, c\right)\stackrel{\mathfrak{u_3}}{\longmapsto}\left(a, c^{-1}\right)$  & $\begin{array}{c}
\mathbf{1}_{r, s}\stackrel{\mathfrak{u_3}}{\longmapsto}\mathbf{1}_{r, -s}\\
\mathbf{3}_{\widetilde{(k, l)}}\stackrel{\mathfrak{u_3}}{\longmapsto}\mathbf{3}_{\widetilde{(-k,-l)}}
\end{array}$   &  $\mathbf{1}_{0, s}\,, \mathbf{3}_{(k, l)}$   &  $\begin{pmatrix}
1  &~  0  &~  0  \\
0  &~  1  &~  0  \\
0  &~  0  &~  1
\end{pmatrix}$  \\
 & & &  \\[-0.14in]\hline
 & & &  \\[-0.14in]
$\left(a, c\right)\stackrel{\mathfrak{u_4}}{\longmapsto}\left(a^2c, c^{-1}\right)$  & $\begin{array}{c}
\mathbf{1}_{r, s}\stackrel{\mathfrak{u_4}}{\longmapsto}\mathbf{1}_{s-r, -s}\\
\mathbf{3}_{\widetilde{(k, l)}}\stackrel{\mathfrak{u_4}}{\longmapsto}\mathbf{3}_{\widetilde{(-l,-k)}}
\end{array}$   &  $\mathbf{1}_{r, 0}\,, \mathbf{3}_{(k, k)}$   &  $\begin{pmatrix}
0  &~  1    &~   0 \\
e^{i\frac{2k\pi}{n}}   &~   0   &~  0 \\
0   &~  0   & ~  e^{i\frac{4k\pi}{n}}
\end{pmatrix}$  \\
 & & &  \\[-0.14in]\hline
 & & &  \\[-0.14in]
$\left(a, c\right)\stackrel{\mathfrak{u_5}}{\longmapsto}\left(a^2c^2, c^{-1}\right)$  & $\begin{array}{c}
\mathbf{1}_{r, s}\stackrel{\mathfrak{u_5}}{\longmapsto}\mathbf{1}_{2s-r, -s}\\
\mathbf{3}_{\widetilde{(k, l)}}\stackrel{\mathfrak{u_5}}{\longmapsto}\mathbf{3}_{\widetilde{(-l,-k)}}
\end{array}$   &  $\mathbf{1}_{r, 0}\,, \mathbf{3}_{(k, k)}$   &  $\begin{pmatrix}
0  &~  1    &~   0 \\
e^{i\frac{4k\pi}{n}}   &~   0   &~  0 \\
0   &~  0   & ~  e^{i\frac{8k\pi}{n}}
\end{pmatrix}$  \\
 & & &  \\[-0.14in]\hline
 & & &  \\[-0.14in]
$\left(a, c\right)\stackrel{\mathfrak{u_6}}{\longmapsto}\left(a^2c, c\right)$  & $\begin{array}{c}
\mathbf{1}_{r, s}\stackrel{\mathfrak{u_6}}{\longmapsto}\mathbf{1}_{s-r, s}\\
\mathbf{3}_{\widetilde{(k, l)}}\stackrel{\mathfrak{u_6}}{\longmapsto}\mathbf{3}_{\widetilde{(l,k)}}
\end{array}$   &  $\mathbf{1}_{r, 0}\,, \mathbf{3}_{(k, n-k)}$   &  $\begin{pmatrix}
0  &~  1    &~   0~ \\
e^{-i\frac{2k\pi}{n}}   &~   0   &~  0~ \\
0   &~  0   & ~  1~
\end{pmatrix}$  \\
 & & &  \\[-0.14in]\hline
 & & &  \\[-0.14in]
$\left(a, c\right)\stackrel{\mathfrak{u_7}}{\longmapsto}\left(a^2c^2, c\right)$  & $\begin{array}{c}
\mathbf{1}_{r, s}\stackrel{\mathfrak{u_7}}{\longmapsto}\mathbf{1}_{2s-r, s}\\
\mathbf{3}_{\widetilde{(k, l)}}\stackrel{\mathfrak{u_7}}{\longmapsto}\mathbf{3}_{\widetilde{(l,k)}}
\end{array}$   &  $\mathbf{1}_{r, 0}\,, \mathbf{3}_{(k, n-k)}$   &  $\begin{pmatrix}
0  &~  1    &~   0 \\
e^{-i\frac{4k\pi}{n}}   &~   0   &~  0 \\
0   &~  0   & ~  1
\end{pmatrix}$  \\
 & & &  \\[-0.14in]\hline
& & &  \\[-0.14in]
$\left(a, c\right)\stackrel{\mathfrak{u_8}}{\longmapsto}\left(ac, c^{-1}\right)$  & $\begin{array}{c}
\mathbf{1}_{r, s}\stackrel{\mathfrak{u_8}}{\longmapsto}\mathbf{1}_{r+s, -s}\\
\mathbf{3}_{\widetilde{(k, l)}}\stackrel{\mathfrak{u_8}}{\longmapsto}\mathbf{3}_{\widetilde{(-k,-l)}}
\end{array}$   &  $\mathbf{1}_{r, r}\,, \mathbf{3}_{(k, l)}$   &  $\begin{pmatrix}
1  &~  0    &~   0 \\
0   &~  e^{-i\frac{2k\pi}{n}}   &~  0 \\
0   &~  0   & ~  e^{i\frac{2l\pi}{n}}
\end{pmatrix}$  \\
 & & &  \\[-0.14in]\hline
& & &  \\[-0.14in]
$\left(a, c\right)\stackrel{\mathfrak{u_9}}{\longmapsto}\left(ac^2, c^{-1}\right)$  & $\begin{array}{c}
\mathbf{1}_{r, s}\stackrel{\mathfrak{u_9}}{\longmapsto}\mathbf{1}_{r+2s, -s}\\
\mathbf{3}_{\widetilde{(k, l)}}\stackrel{\mathfrak{u_9}}{\longmapsto}\mathbf{3}_{\widetilde{(-k,-l)}}
\end{array}$   &  $\begin{array}{l}
\mathbf{1}_{0, 0}\,, \mathbf{1}_{1, 2},\\
\mathbf{1}_{2, 1}\,, \mathbf{3}_{(k, l)}
\end{array}$   &  $\begin{pmatrix}
1  &~  0    &~   0 \\
0   &~  e^{-i\frac{4k\pi}{n}}   &~  0 \\
0   &~  0   & ~  e^{i\frac{4l\pi}{n}}
\end{pmatrix}$  \\
 & & &  \\[-0.14in]\hline
& & &  \\[-0.14in]
$\left(a, c\right)\stackrel{\mathfrak{u_{10}}}{\longmapsto}\left(ac, c\right)$  & $\begin{array}{c}
\mathbf{1}_{r, s}\stackrel{\mathfrak{u_{10}}}{\longmapsto}\mathbf{1}_{r+s, s}\\
\mathbf{3}_{\widetilde{(k, l)}}\stackrel{\mathfrak{u_{10}}}{\longmapsto}\mathbf{3}_{\widetilde{(k,l)}}
\end{array}$   &  $\mathbf{1}_{0, 0}\,, \mathbf{3}_{(\frac{n}{2}, \frac{n}{2})}$   &  $\begin{pmatrix}
1  &~  0    &~   0 \\
0   &~  -1   &~  0 \\
0   &~  0   & ~  -1
\end{pmatrix}$  \\
 & & &  \\[-0.14in]\hline
& & &  \\[-0.14in]
$\left(a, c\right)\stackrel{\mathfrak{u_{11}}}{\longmapsto}\left(ac^2, c\right)$  & $\begin{array}{c}
\mathbf{1}_{r, s}\stackrel{\mathfrak{u_{11}}}{\longmapsto}\mathbf{1}_{r+2s, s}\\
\mathbf{3}_{\widetilde{(k, l)}}\stackrel{\mathfrak{u_{11}}}{\longmapsto}\mathbf{3}_{\widetilde{(k,l)}}
\end{array}$   &  $\mathbf{1}_{0, 0}\,, \mathbf{3}_{(\frac{n}{2}, \frac{n}{2})}$   &  $\begin{pmatrix}
1  &~  0    &~   0 \\
0   &~  1   &~  0 \\
0   &~  0   & ~  1
\end{pmatrix}$  \\
 & & &  \\[-0.14in]\hline\hline
\end{tabular}
\caption{\label{tab:GCP_n=3N}Classification of the automorphisms of the $\Delta(3n^2)$ group with $n=3\, \mathbb{Z}$ and the corresponding CP transformations, where the parameters $r, s=0, 1, 2$ and $k, l=0, 1,\ldots, n-1$. Note that the triplet representation $\mathbf{3}_{(k, l)}$ with $(k, l)=(0, 0)$, $(\frac{n}{3}, \frac{n}{3})$, $(\frac{2n}{3}, \frac{2n}{3})$ is reducible. ``rep'' is the abbreviation of ``representation''. ``Rep contents'' stands for the irreducible representations which can be present in a model if one impose the CP transformation corresponding to a automorphism to be a symmetry. }
\end{center}
\end{table}

\section{\label{sec:mixing_semi-direct}Possible lepton mixing from the semi-direct approach}

In the present work, both family symmetry $\Delta(3n^2)$ and the generalized CP symmetry would be included at high energy scale. We shall perform a model-independent analysis of the possible lepton mixing patterns enforced by the residual symmetries of the neutrino and the charged lepton sectors. As usual, the three generations of the lepton doublet fields are assigned to an irreducible triplet representation of $\Delta(3n^2)$. The light neutrinos are assumed to be Majorana particles such that the residual flavor symmetry $G_{\nu}$ of the neutrino sector must be restricted to a Klein subgroup or a $Z_2$ subgroup of $\Delta(3n^2)$.

$\Delta(3n^2)$ has a unique Klein subgroup $\left\{1, c^{n/2}, d^{n/2}, c^{n/2}d^{n/2}\right\}$ if and only if $n$ is an even number. In the case that the full Klein symmetry is preserved by the neutrino mass matrix, i.e., $G_{\nu}=\left\{1, c^{n/2}, d^{n/2}, c^{n/2}d^{n/2}\right\}$, and the residual flavor symmetry $G_{l}$ of the charged lepton sector ia an abelian subgroup generated by the generators $c$ and $d$, the lepton mixing matrix would be a trivial unit matrix up to permutations of rows and columns. Obviously this scenario is already excluded by the present neutrino oscillation data. On the other hand, if $G_{\nu}=\left\{1, c^{n/2}, d^{n/2}, c^{n/2}d^{n/2}\right\}$ and $G_{l}$ is a $Z_3$ subgroup with $G_{l}=Z^{(s,t)}_3\equiv\left\{1, ac^sd^t, a^2c^{s-t}d^{s}\right\}$, we find that all elements of the PMNS matrix have the same absolute value, i.e.
\begin{equation}
||U_{\rm{PMNS}}||=\frac{1}{\sqrt{3}}\begin{pmatrix}
1  &~  1  &~  1 \\
1  &~  1  &~  1 \\
1  &~  1  &~  1
\end{pmatrix}\,,
\end{equation}
Therefore both solar mixing angles and atmospheric mixing angles are predicted to be maximal, and the value of the reactor mixing angle is $\sin^2\theta_{13}=1/3$ which is outside the experimentally preferred $3\sigma$ ranges~\cite{Gonzalez-Garcia:2014bfa,Capozzi:2013csa,Forero:2014bxa}.  In the next step, we shall proceed to analyze all the possible lepton mixing patterns in the ``semi-direct'' method. In this approach, the $\Delta(3n^2)$ flavor symmetry is broken down to an abelian subgroup $G_{l}$ in the charged lepton sector, and the remnant symmetry preserved by the neutrino mass matrix is $Z_2\times CP$. If $G_{l}$ is generated by $c$ and $d$, one column of the PMNS matrix would be fixed to be $\left(1,0,0\right)^{T}$ up to permutations. The experimental data can not be accommodated. As a result, we shall concentrate on the case of $G_{l}=Z^{(s,t)}_3\equiv\left\{1, ac^sd^t, a^2c^{s-t}d^{s}\right\}$ with $s, t=0, 1, \ldots, n-1$.
The three generations of the left-handed lepton doublets are assumed to transform as an irreducible three-dimensional representation $\mathbf{3}_{(k, l)}$ of $\Delta(3n^2)$. The invariance of the charged lepton mass matrix $m_{l}$ under the residual flavor symmetry  $G_{l}=Z^{(s,t)}_3\equiv\left\{1, ac^sd^t, a^2c^{s-t}d^{s}\right\}$ leads to
\begin{equation}
\label{eq:invariace_rem_flav}\rho^{\dagger}_{\mathbf{3}_{(k, l)}}(ac^sd^t)m^{\dagger}_{l}m_{l}\rho_{\mathbf{3}_{(k, l)}}(ac^sd^t)=m^{\dagger}_{l}m_{l}\,.
\end{equation}
This implies that $m^{\dagger}_{l}m_{l}$ and $\rho_{\mathbf{3}_{(k, l)}}(ac^sd^t)$ are commutable with each other and they can be diagonalized by the same unitary matrix. From the explicit from of the representation matrices given in Eq.~\eqref{eq:3-dim_rep}, we find that
\begin{equation}
U^{\dagger}_{l}\rho_{\mathbf{3}_{(k, l)}}(ac^sd^t)U_{l}=\text{diag}\left(1, \omega, \omega^2\right)\,,
\end{equation}
where the unitary transformation $U_{l}$ is
\begin{equation}
U_{l}=\frac{1}{\sqrt{3}}\begin{pmatrix}
e^{-2\pi i\frac{ls-(k+l)t}{n}}   &~   \omega e^{-2\pi i\frac{ls-(k+l)t}{n}}  &~  \omega^2e^{-2\pi i\frac{ls-(k+l)t}{n}}\\
e^{-2\pi i\frac{(k+l)s-kt}{n}}   &~   \omega^2e^{-2\pi i\frac{(k+l)s-kt}{n}}  &~  \omega e^{-2\pi i\frac{(k+l)s-kt}{n}} \\
1   &~   1    &~  1
\end{pmatrix}\,.
\end{equation}
From the invariance condition of Eq.~\eqref{eq:invariace_rem_flav}, it follows that $U_{l}$ also diagonalizes $m^{\dagger}_{l}m_{l}$. Then we turn to the neutrino sector. The $\Delta(3n^2)$ flavor symmetry has three $Z_2$ subgroups $\left\{1, c^{n/2}\right\}$, $\left\{1, d^{n/2}\right\}$ and $\left\{1, c^{n/2}d^{n/2}\right\}$ if $n$ is divisible by 2. Since different pair of residual symmetries related by similarity transformation lead to the same predictions for the lepton flavor mixing. Without loss of generality, the residual $Z_2$ flavor symmetry of the neutrino sector can be chosen to be $G_{\nu}=\left\{1, c^{n/2}\right\}$. The residual CP transformation $X_{\nu}$ in the neutrino sector should be compatible with the residual flavor symmetry symmetry $G_{\nu}=\left\{1, c^{n/2}\right\}$, and the following consistency equation must be satisfied
\begin{equation}
\label{eq:consistency_condition_neutrino}X_{\nu}\rho^{\ast}_{\mathbf{3}_{(k, l)}}(c^{n/2})X^{\dagger}_{\nu}=\rho_{\mathbf{3}_{(k, l)}}(c^{n/2})\,.
\end{equation}
Moreover, $X_{\nu}$ should be a symmetric unitary matrix otherwise the light neutrino mass would be partially degenerate~\cite{Feruglio:2012cw,Chen:2014wxa}. In the following, we shall investigate the remnant CP transformation $X_{\nu}$ for the generalized CP transformation defined by the automorphisms $\mathfrak{u}_1$, $\mathfrak{u}_2$ and $\mathfrak{u}_3$ one by one, and the predictions for the PMNS matrix and the lepton mixing parameters would be presented.

Firstly we impose the generalized CP symmetry corresponding to the automorphism $\mathfrak{u}_1$ onto the model. Then only the three-dimensional representation $\mathbf{3}_{(k, k)}$ can be present. The left-handed lepton doublets are embedded into a triplet $\mathbf{3}_{(1,1)}$ which form a faithful representation of $\Delta(6n^2)$ if $n$ is not divisible by $3$. Solving the consistency equation of Eq.~\eqref{eq:consistency_condition_neutrino}, we find that the residual CP transformation $X_{\nu}$ is $\rho_{\mathbf{3}_{(1,1)}}(c^{\gamma}d^{zn/3})X_{\mathfrak{u}_1}$, where $\gamma=0, 1, \ldots, n-1$, and $z=0$ for $n\neq3\, \mathbb{Z}$ and $z=0$, $1$, $2$ for $n=3\, \mathbb{Z}$. The invariance of the neutrino mass matrix $m_{\nu}$ under the remnant flavor symmetry $\rho_{\mathbf{3}_{(1,1)}}(c^{n/2})$ and the residual CP transformation $X_{\nu}$ implies
\begin{equation}
\rho^{T}_{\mathbf{3}_{(1,1)}}(c^{n/2})m_{\nu}\rho_{\mathbf{3}_{(1,1)}}(c^{n/2})=m_{\nu},\qquad X^{T}_{\nu}m_{\nu}X_{\nu}=m^{\ast}_{\nu}\,.
\end{equation}
Then $m_{\nu}$ is fixed to be
\begin{equation}
m_{\nu}=\omega^{-z}\begin{pmatrix}
e^{i\theta}m_{11}   &~  \eta^{-\gamma}m_{12}  &~   0  \\
\eta^{-\gamma}m_{12}    &~   \eta^{-2\gamma}e^{-i\theta}m_{11}    & ~  0  \\
0   & ~  0   & ~ \eta^{2\gamma}m_{33}
\end{pmatrix}\,,
\end{equation}
where $\theta$, $m_{11}$, $m_{12}$ and $m_{33}$ are real parameters. The neutrino mass matrix $m_{\nu}$ is diagonalized by the following unitary matrix
\begin{equation}
U_{\nu}=\frac{\omega^{-z}}{\sqrt{2}}\begin{pmatrix}
e^{-i\frac{\theta}{2}}    &~    e^{-i\frac{\theta}{2}}   &~   0  \\
-e^{i(\frac{\theta}{2}+\frac{2\gamma\pi}{n})}   &~  e^{i(\frac{\theta}{2}+\frac{2\gamma\pi}{n})}   &~  0  \\
0    &~   0   & ~  \sqrt{2}\,e^{-i\frac{2\gamma\pi}{n}}
\end{pmatrix}\,,
\end{equation}
with
\begin{equation}
U^{T}_{\nu}m_{\nu}U_{\nu}=\text{diag}\left(m_{11}-m_{12}, m_{11}+m_{12}, m_{33}\right)\,.
\end{equation}
Therefore the lepton flavor mixing matrix is constrained by the remnant symmetry to be
\begin{equation}
\label{eq:PMNS_u1}U_{PMNS}=U^{\dagger}_{l}U_{\nu}=\frac{1}{\sqrt{6}}\begin{pmatrix}
1-e^{i\varphi}   &~  \sqrt{2}    &~   1+e^{i\varphi}  \\
\omega^2-\omega e^{i\varphi}   &~  \sqrt{2}    & ~  \omega^2+\omega e^{i\varphi}  \\
\omega-\omega^2 e^{i\varphi}   &~  \sqrt{2}    & ~  \omega+\omega^2 e^{i\varphi}
\end{pmatrix}\text{diag}(e^{i\rho},1, e^{i\rho})Q_{\nu}\,,
\end{equation}
where $Q_{\nu}$ is a diagonal matrix with non-vanishing entry $\pm1$ or $\pm i$, it sets the light neutrino masses being positive, and it can shift the Majorana phases by $\pi$. The parameters $\varphi$ and $\rho$ are given by
\begin{equation}
\varphi=\theta+\frac{2\pi(s+t+\gamma)}{n},\qquad  \rho=-\frac{\theta}{2}+\frac{2\pi(s-2t+\gamma)}{n} \,.
\end{equation}
We see that the second column of the PMNS matrix is $(1, 1, 1)^{T}/\sqrt{3}$, and therefore it is trimaximal mixing. The lepton mixing parameters read as
\begin{eqnarray}
\nonumber&&\sin^2\theta_{13}=\frac{1}{3}(1+\cos\varphi),\qquad \sin^2\theta_{12}=\frac{1}{2-\cos\varphi},\qquad \sin^2\theta_{23}=\frac{1}{2}+\frac{\sqrt{3}\sin\varphi}{2(2-\cos\varphi)},\\
\label{eq:mixing_para_u1}&&\sin\delta_{CP}=0,\quad \tan\alpha_{21}=-\tan(2\rho+\varphi)=-\tan\frac{6\pi(s-t+\gamma)}{n},\quad \sin\alpha_{31}=0\,.
\end{eqnarray}
Therefore both Dirac CP phase $\delta_{CP}$ and one of the Majorana phase $\alpha_{31}$ are conserved, and the value of another Majorana phase $\alpha_{21}$ is equal to $-6\pi(s-t+\gamma)/n$ or $\pi-6\pi(s-t+\gamma)/n$. Hence $\alpha_{21}\hskip-0.06in\pmod{2\pi}$ can take the discrete values $0$, $\frac{2\pi}{n}$, $\ldots$, $2\pi-\frac{2\pi}{n}$ for $n \neq 3\, \mathbb{Z}$, and the admissible values of $\alpha_{21}\hskip-0.06in\pmod{2\pi}$ are $0$, $\frac{2\pi}{n^{\prime}}$, $\ldots$, $2\pi-\frac{2\pi}{n^{\prime}}$ if $n$ is divisible by 3 with $n=3n^{\prime}$. Since all the three mixing angles $\theta_{12}$, $\theta_{13}$ and $\theta_{23}$ depends on a single common parameter $\varphi$, they are strongly correlated as follows
\begin{equation}
\label{eq:correlation_u1}3\sin^2\theta_{12}\cos^2\theta_{13}=1,\qquad \sin^2\theta_{23}=\frac{1}{2}\pm\frac{1}{2}\tan\theta_{13}\sqrt{2-\tan^2\theta_{13}}\,. \end{equation}
Given the precisely measured reactor mixing angle $1.77\times10^{-2}\leq\sin^2\theta_{13}\leq2.94\times10^{-2}$~\cite{Forero:2014bxa}, we have
\begin{equation}
0.339\leq\sin^2\theta_{12}\leq0.343,\quad 0.378\leq\sin^2\theta_{23}\leq0.406~~\mathrm{or}~~0.594\leq\sin^2\theta_{23}\leq0.622\,.
\end{equation}
These predictions can be tested in near future neutrino oscillation experiments. Subsequently we proceed to consider the generalized CP symmetry corresponding to the automorphism $\mathfrak{u}_2$. The model can only contain the triplet representations $\mathbf{3}_{k, n-k}$. The left-handed lepton doublets are assigned to a three-dimensional representation $\mathbf{3}_{(1, n-1)}$ which always gives rise to a faithful representation of $\Delta(3n^2)$. The residual CP transformation is determined to be $X_{\nu}=\rho_{\mathbf{3}_{(1,n-1)}}(c^{\gamma}d^{2\gamma})X_{\mathfrak{u}_2}$ with $\gamma=0, 1, \ldots, n-1$. The light neutrino mass matrix is constrained by the remnant symmetry to be of the following form
\begin{equation}
m_{\nu}=\begin{pmatrix}
e^{i\theta}m_{11}   &~   \eta^{\gamma}m_{12}   &~  0  \\
\eta^{\gamma}m_{12}   & ~  \eta^{2\gamma}e^{-i\theta}m_{11}   &~   0  \\
0   &~  0  & ~  \eta^{-2\gamma}m_{33}
\end{pmatrix}\,,
\end{equation}
where $m_{11}$, $m_{12}$, $m_{33}$ and $\theta$ are real. The unitary transformation $U_{\nu}$ reads
\begin{equation}
U_{\nu}=\frac{1}{\sqrt{2}}\begin{pmatrix}
e^{-i\frac{\theta}{2}}    &~   e^{-i\frac{\theta}{2}}    & ~  0  \\
-e^{i(\frac{\theta}{2}-\frac{2\gamma\pi}{n})}   & ~  e^{i(\frac{\theta}{2}-\frac{2\gamma\pi}{n})}    &~  0 \\
0    & ~  0   & ~   \sqrt{2}\,e^{i\frac{2\gamma\pi}{n}}
\end{pmatrix}\,.
\end{equation}
Consequently the PMNS matrix is of the same form as Eq.~\eqref{eq:PMNS_u1}, and the parameters $\varphi$ and $\rho$ are
\begin{equation}
\varphi=\theta+\frac{2\pi(s-t-\gamma)}{n},\qquad  \rho=-\frac{\theta}{2}-\frac{2\pi(s+\gamma)}{n} \,.
\end{equation}
The predictions for the lepton mixing parameters are given in Eq.~\eqref{eq:mixing_para_u1}. The measured value of the reactor angle $\theta_{13}$ can be achieved for appropriate values of $\theta$. The correlations among the mixing angles are shown in Eq.~\eqref{eq:correlation_u1}. The analytical expression for the Majorana phase $\alpha_{21}$ is $2\pi(s+t+3\gamma)/n$ or $\pi+2\pi(s+t+3\gamma)/n$. Hence $\alpha_{21}$ can take the discrete values $0$, $2\pi/n$, $4\pi/n$, $\ldots$ and $2\pi-2\pi/n$ no matter whether $n$ can be divisible by 3 or not. Note that this type of remnant symmetry also appears in the context of $\Delta(6n^2)$ family symmetry with generalized CP~\cite{Hagedorn:2014wha,Ding:2014ora}, and it corresponds to the case IV of~\cite{Ding:2014ora}. We have checked that the same results are obtained here.

\begin{figure}[t!]
\begin{center}
\includegraphics[width=0.48\textwidth]{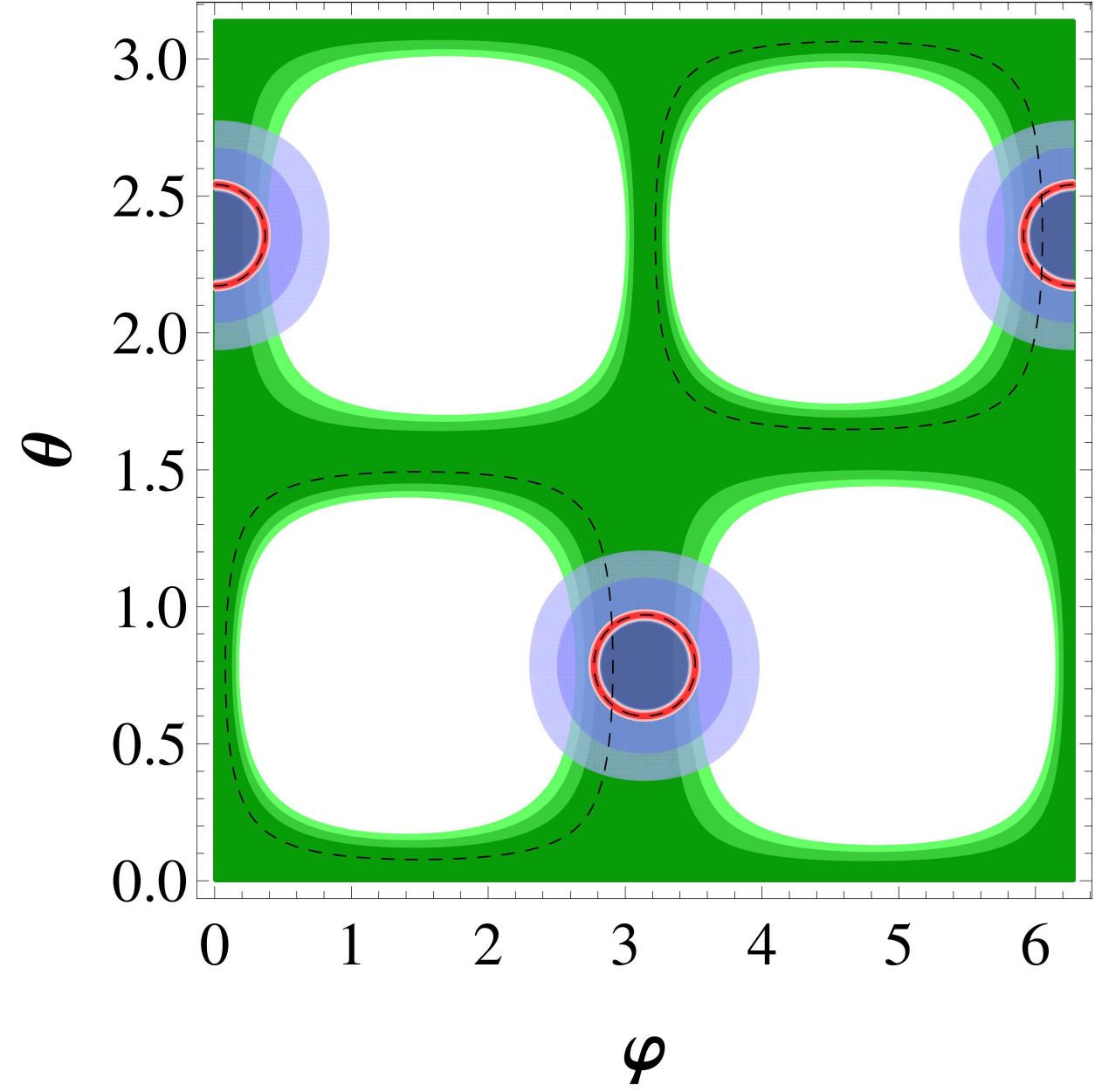}
\includegraphics[width=0.48\textwidth]{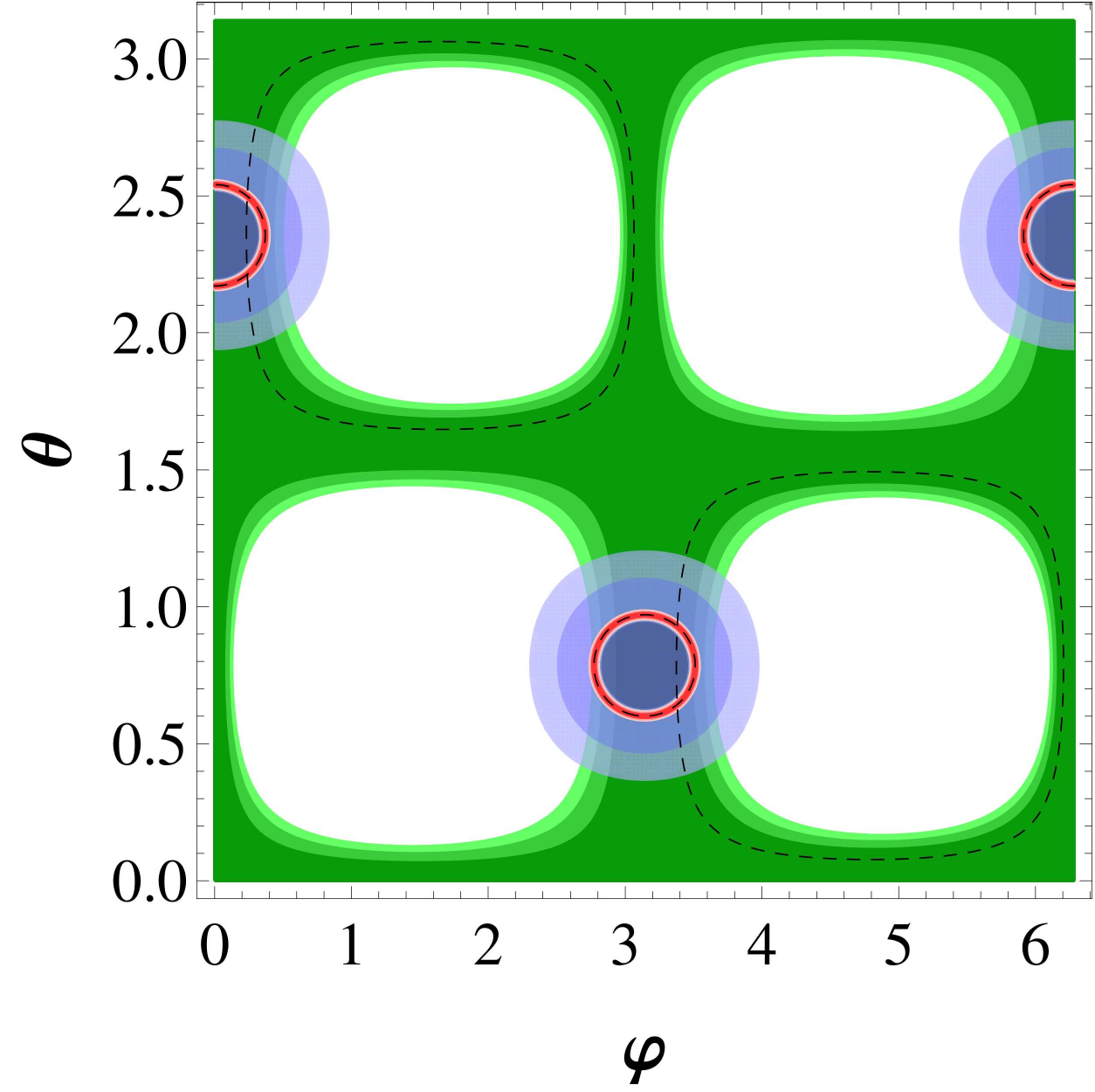} \caption{\label{fig:contour_angles}The blue, red and green regions are the phenomenologically allowed regions for $\sin^2\theta_{12}$, $\sin^2\theta_{13}$ and  $\sin^2\theta_{23}$ respectively in the plane of $\theta$ and $\varphi$, according to Eq.~\eqref{eq:mixing_para_u3}. The allowed $1\sigma$, $2\sigma$ and $3\sigma$ ranges of the lepton mixing angles are denoted by different shadings of each colour, while the dashed lines denote the best fitting contours of $\sin^2\theta_{13}$ and $\sin^2\theta_{23}$. Note that the best fitting value of $\sin^2\theta_{12}$ can not be achieved since $\sin^2\theta_{12}$ is bound from below with $\sin^2\theta_{12}>1/3$ in this case. Here we use the data from the global fit in~\cite{Forero:2014bxa}. The two panels differ in the permutation of the second and the third row of the the PMNS matrix.}
\end{center}
\end{figure}

\begin{figure}[t!]
\begin{center}
\includegraphics[width=0.495\textwidth]{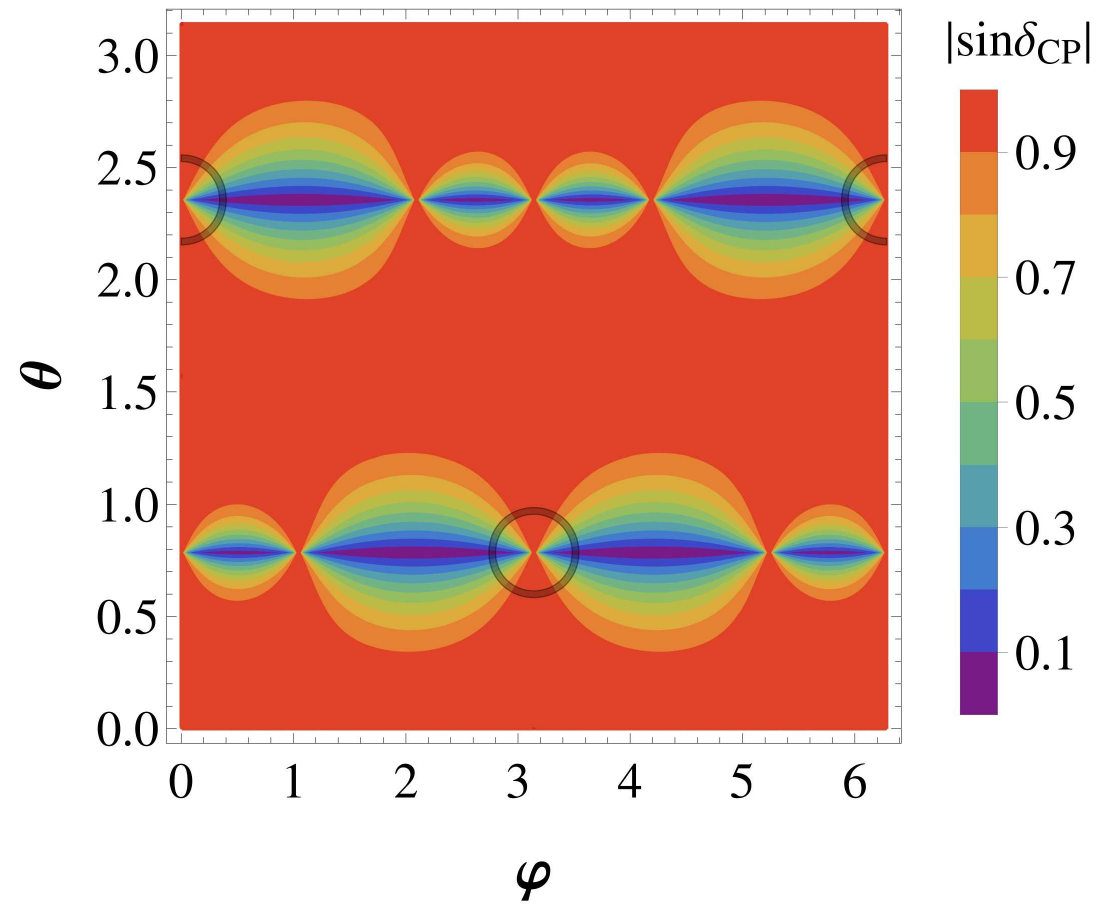}
\includegraphics[width=0.495\textwidth]{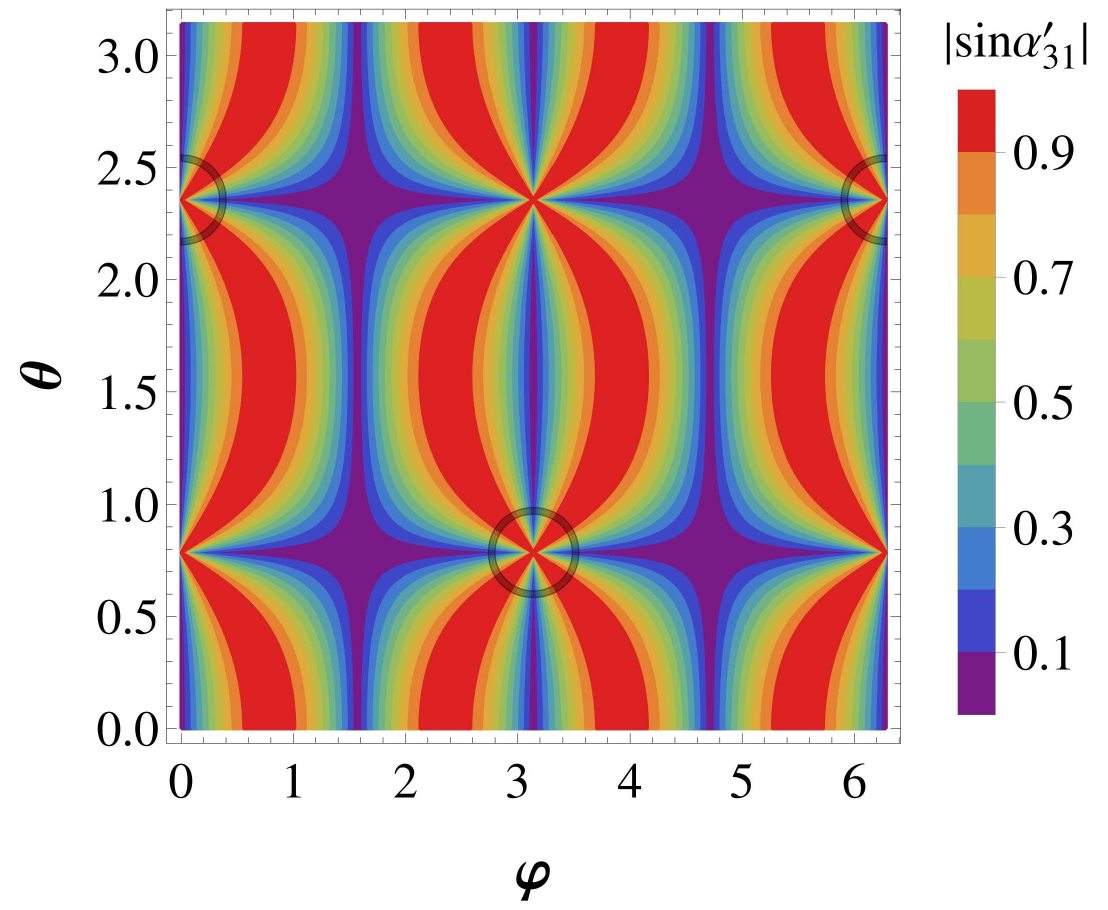} \caption{\label{fig:contour_CP}
The contour plots for the predictions of $|\sin\delta_{CP}|$ and $|\sin\alpha^{\prime}_{31}|$,
according to Eq.~\eqref{eq:mixing_para_u3}.
The black areas denote the regions of $\theta$ and $\varphi$ for which the three lepton mixing angles are compatible with the experimental data at $3\sigma$ level.}
\end{center}
\end{figure}

Now we turn to the third kind of generalized CP transformation defined by the automorphism $\mathfrak{u}_3$. It maps each three-dimensional representation into its complex conjugate representation such that all three-dimensional representations can be present. The lepton doublet fields are assumed to transform as $\mathbf{3}_{(1, n-1)}$ which is a faithful representation of $\Delta(3n^2)$ for any value of $n$. The $Z_2$ residual flavor symmetry $G_{\nu}=\left\{1, c^{n/2}\right\}$ can be combined with the CP transformation $X_{\nu}=\rho_{\mathbf{3}_{(1, n-1)}}(c^{\gamma}d^{\delta})$. This case has been discussed in~\cite{Hagedorn:2014wha}, it is also permissible in $\Delta(6n^2)$ and generalized CP~\cite{Hagedorn:2014wha,Ding:2014ora}, and it is exactly the case III of~\cite{Ding:2014ora}. The light neutrino mass matrix can be derived from the assumed residual symmetries as follows
\begin{equation}
m_{\nu}=\begin{pmatrix}
\eta^{\gamma}m_{11}   &~   \eta^{\frac{\delta}{2}}m_{12}   &~   0  \\
\eta^{\frac{\delta}{2}}m_{12}   &~   \eta^{-\gamma+\delta}m_{22}   &~   0 \\
0   &~   0   &~   \eta^{-\delta}m_{33}
\end{pmatrix}\,,
\end{equation}
where $m_{11}$, $m_{22}$, $m_{23}$ and $m_{33}$ are real. The unitary transformation $U_{\nu}$ which diagonalizes this  neutrino mass matrix is given by
\begin{equation}
U_{\nu}=\begin{pmatrix}
e^{-i\pi\frac{\gamma}{n}}\cos\theta    &~   e^{-i\pi\frac{\gamma}{n}}\sin\theta    &~   0   \\
-e^{i\pi\frac{\gamma-\delta}{n}}\sin\theta    &~  e^{i\pi\frac{\gamma-\delta}{n}}\cos\theta   &~   0   \\
0   &~   0   &~  e^{i\pi\frac{\delta}{n}}
\end{pmatrix}\,,
\end{equation}
with
\begin{equation}
\tan2\theta=\frac{2m_{12}}{m_{22}-m_{11}}\,.
\end{equation}
The light neutrino mass eigenvalues are
\begin{equation}
m_{1}=\frac{1}{2}(m_{11}+m_{22})-\frac{m_{22}-m_{11}}{2\cos2\theta},\quad m_{2}=\frac{1}{2}(m_{11}+m_{22})+\frac{m_{22}-m_{11}}{2\cos2\theta},\quad m_{3}=m_{33},\quad \,.
\end{equation}
Then the PMNS matrix is still of the trimaximal form
\begin{equation}
\label{eq:PMNS_u3}U_{PMNS}=\frac{1}{\sqrt{3}}\begin{pmatrix}
\cos\theta-e^{i\varphi}\sin\theta   &~  1   &~  \sin\theta+e^{i\varphi}\cos\theta \\
\omega^2\cos\theta-\omega e^{i\varphi}\sin\theta   & ~ 1   &~  \omega^2\sin\theta+\omega e^{i\varphi}\cos\theta  \\
\omega\cos\theta-\omega^2e^{i\varphi}\sin\theta   &~  1   &~  \omega\sin\theta+\omega^2e^{i\varphi}\cos\theta
\end{pmatrix}\text{diag}(e^{i\rho}, 1, e^{i\rho})\,,
\end{equation}
where
\begin{equation}
\varphi=\frac{\pi(2s-2t+2\gamma-\delta)}{n},\qquad \rho=-\frac{\pi(2 s+\gamma+\delta)}{n}\,,
\end{equation}
which can take the values
\begin{equation}
\varphi, \rho\hskip-0.08in\pmod{2\pi}=0, \frac{\pi}{n},\ldots,2\pi-\frac{\pi}{n}\,. \end{equation}
The lepton mixing parameters are determined to be
\begin{eqnarray}
\nonumber&&\sin^2\theta_{13}=\frac{1}{3}(1+\cos\varphi\sin2\theta),\qquad \sin^2\theta_{12}=\frac{1}{2-\cos\varphi\sin2\theta},\\
\nonumber&&\sin^2\theta_{23}=\frac{1-\cos(\varphi+\frac{\pi}{3})\sin2\theta}{2-\cos\varphi\sin2\theta},\qquad \tan\delta_{CP}=-\frac{\left(2-\cos\varphi\sin2\theta\right)\cot2\theta}{\left(1-2\cos\varphi\sin2\theta\right)\sin\varphi}\,,\\
\nonumber&&\tan\alpha_{21}=-\frac{\cos^2\theta\sin2\rho+\sin^2\theta\sin2\left(\rho+\varphi\right)-\sin2\theta\sin\left(2\rho+\varphi\right)}
{\cos^2\theta\cos2\rho+\sin^2\theta\cos2\left(\rho+\varphi\right)-\sin2\theta\cos\left(2\rho+\varphi\right)}\,,\\
\label{eq:mixing_para_u3}&&\tan\alpha^{\prime}_{31}=\frac{4\cos2\theta\sin2\varphi}{-1+3\cos2\varphi+2\cos^2\varphi\cos4\theta}
\,,
\end{eqnarray}
where $\alpha^{\prime}_{31}=\alpha_{31}-2\delta_{CP}$. The solar angle and reactor angle are related by $3\sin^2\theta_{12}\cos^2\theta_{13}=1$ such that $\sin^2\theta_{12}>1/3$ is fulfilled. Notice that the PMNS matrix is determined up to possible permutations of rows and columns since both the neutrino and charged lepton mass orders can not be fixed by residual symmetry. If the second and third rows of the PMNS matrix in Eq.~\eqref{eq:PMNS_u3} are exchanged, $\theta_{23}$ and $\delta_{CP}$ become $\pi/2-\theta_{23}$ and $\pi+\delta_{CP}$ respectively while the remaining mixing parameters keep intact. The three CP rephasing invariants $J_{CP}$, $I_1$ and $I_2$ are given by
\begin{eqnarray}
\nonumber&&\qquad\qquad\quad ~J_{CP}=\frac{1}{6\sqrt{3}}\cos2\theta,\qquad I_2=\frac{1}{9}\cos2\theta\sin2\varphi,\\
\label{eq:invariants_u3}&&I_1=-\frac{1}{9}\left[\cos^2\theta\sin2\rho+\sin^2\theta\sin2\left(\rho+\varphi\right)-\sin2\theta\sin\left(2\rho+\varphi\right)\right]\,,
\end{eqnarray}
where $J_{CP}$ is the Jarlskog invariant~\cite{Jarlskog:1985ht},
\begin{eqnarray}
\nonumber J_{CP}&=&\text{Im}\left[\left(U_{PMNS}\right)_{11}\left(U_{PMNS}\right)_{33}\left(U^{\ast}_{PMNS}\right)_{13}\left(U^{\ast}_{PMNS}\right)_{31}\right]\\
&=&\frac{1}{8}\sin2\theta_{12}\sin2\theta_{13}\sin2\theta_{23}\cos\theta_{13}\sin\delta_{CP}\,.
\end{eqnarray}
The remaining two invariants $I_1$ and $I_2$ are related with the Majorana phases and they are defined as~\cite{Branco:1986gr,Jenkins:2007ip,Branco:2011zb},
\begin{eqnarray}
\nonumber I_1&=&\text{Im}\left[\left(U_{PMNS}\right)^2_{12}\left(U_{PMNS}\right)^{\ast2}_{11}\right]=\frac{1}{4}\sin^22\theta_{12}\cos^4\theta_{13}\sin\alpha_{21}\,,\\
I_2&=&\text{Im}\left[\left(U_{PMNS}\right)^2_{13}\left(U_{PMNS}\right)^{\ast2}_{11}\right]=\frac{1}{4}\sin^22\theta_{13}\cos^2\theta_{12}\sin\alpha^{\prime}_{31}\,,
\end{eqnarray}
Moreover, we see that the Majorana phase $\alpha_{21}$ depends on $\rho$, $\varphi$ and the continuous parameter $\theta$ while all the three mixing angles $\theta_{12}$, $\theta_{13}$, $\theta_{23}$ and the remaining CP phases $\delta_{CP}$, $\alpha^{\prime}_{31}$ only depend on $\varphi$ and $\theta$. The permissible regions of $\varphi$ and $\theta$ allowed by the measured values of the mixing angles $\sin^2\theta_{13}$, $\sin^2\theta_{12}$ and $\sin^2\theta_{23}$ at $3\sigma$ level are displayed in Fig.~\ref{fig:contour_angles}. It is remarkable that the best fit values of $\theta_{13}$ and $\theta_{23}$ can be achieved simultaneously, and the tightest constraint arises from the reactor angle $\theta_{13}$. The relative smallness of $\theta_{13}$ leads to
\begin{equation}
(\varphi, \theta)\simeq(0, 3\pi/4), ~(\pi, \pi/4),~ (2\pi, 3\pi/4)\,.
\end{equation}
The contour plots for $|\sin\delta_{CP}|$ and $|\sin\alpha^{\prime}_{31}|$ are shown in Fig.~\ref{fig:contour_CP}. Since the sign of $\sin\delta_{CP}$ and $\sin\alpha^{\prime}_{31}$ would be reversed if the lepton fields are assigned to $\mathbf{3}_{n-1, 1}$ instead, their absolute values are presented. Notice that almost any value of $\delta_{CP}$ can be achieved in the regions where all three mixing angles are within the experimentally favored $3\sigma$ range.

In the same fashion, we can study the phenomenological predictions of the remaining generalized CP listed in Table~\ref{tab:GCP_n=3N} in the case of $n=3\, \mathbb{Z}$. For the CP transformations corresponding to the automorphisms $\mathfrak{u}_4$ and $\mathfrak{u}_5$, only the irreducible representations $\mathbf{1}_{r, 0}$ and $\mathbf{3}_{(k, k)}$ with $r\neq0, n/3, 2n/3$ can be present. However, we can not find a symmetric remnant CP transformation $X_{\nu}$ such that the restricted consistency condition of Eq.~\eqref{eq:consistency_condition_neutrino} is fulfilled for $k=l=1$. As a result, the lepton mixing is constrained by the remnant flavor symmetry to be the trimaximal pattern but no predictions on the three CP phases $\delta_{CP}$, $\alpha_{21}$ and $\alpha_{31}$ can be extracted~\cite{Petcov:2014laa,Girardi:2015rwa}.

The CP transformations corresponding to $\mathfrak{u}_6$ and $\mathfrak{u}_7$ can be consistently defined if a model only contains the representations $\mathbf{1}_{r, 0}$ and $\mathbf{3}_{(k, n-k)}$ with $k\neq0$. The residual CP transformation is $X_{\nu}=\rho_{\mathbf{3}_{(k, n-k)}}(c^{\gamma}d^{2\gamma-1})X_{\mathfrak{u}_6}$ and $X_{\nu}=\rho_{\mathbf{3}_{(k, n-k)}}(c^{\gamma}d^{2\gamma-2})X_{\mathfrak{u}_7}$ respectively. The three lepton doublet fields are assumed to transform as $\mathbf{3}_{(1, n-1)}$. The PMNS matrix is found to be of the same form as Eq.~\eqref{eq:PMNS_u1}, and the parameters $\rho$ and $\varphi$ are
\begin{eqnarray}
\nonumber\varphi=\theta+\frac{2\pi(s-t-\gamma)}{n},\quad \rho=-\frac{\theta}{2}+\frac{\pi(1-2s-2\gamma)}{n},&& \quad  \text{for}~ \mathfrak{u}_6\,,\\
\varphi=\theta+\frac{2\pi(s-t-\gamma)}{n},\quad \rho=-\frac{\theta}{2}+\frac{\pi(2-2s-2\gamma)}{n}, && \quad  \text{for}~ \mathfrak{u}_7\,.
\end{eqnarray}
As a result, the lepton mixing parameters are still given by Eq.~\eqref{eq:mixing_para_u1}. The Majorana phase $\alpha_{21}$ is equal to
$2\pi(-1+s+t+3\gamma)/n$ and $2\pi(-2+s+t+3\gamma)/n$ up to $\pi$ respectively. It can take the values of $0$, $2\pi/n$, \ldots, $2\pi-2\pi/n$. We conclude that the same predictions for the lepton mixing as the case of $\mathfrak{u}_2$ are obtained.

For the CP transformations associated with the automorphisms $\mathfrak{u}_{8}$ and $\mathfrak{u}_{9}$, any triplet representation can be present. The remnant CP transformation compatible with the residual flavor symmetry $G_{\nu}=\left\{1, c^{n/2}\right\}$ is $X_{\nu}=\rho_{\mathbf{3}_{(k, l)}}(c^{\gamma}d^{\delta})X_{\mathfrak{u}_8}$ and $X_{\nu}=\rho_{\mathbf{3}_{(k, l)}}(c^{\gamma}d^{\delta})X_{\mathfrak{u}_9}$ respectively. The left-handed lepton doublet fields are embedded into a faithful three-dimensional
representation $\mathbf{3}_{(1,n-1)}$. We find that the lepton mixing matrix takes exactly the same form as Eq.~\eqref{eq:PMNS_u3} with
\begin{eqnarray}
\nonumber\varphi=\frac{\pi(2s-2t+2\gamma-\delta-1)}{n},\quad \rho=-\frac{\pi  (2s+\gamma+\delta-1)}{n},&& \quad  \text{for}~ \mathfrak{u}_8\,,\\
\varphi=\frac{\pi(2s-2t+2\gamma-\delta-2)}{n},\quad \rho=-\frac{\pi(2s+\gamma+\delta-2)}{n},&& \quad  \text{for}~ \mathfrak{u}_9\,.
\end{eqnarray}
The analytical expressions for the lepton mixing parameters and rephasing invariants are still given by Eq.~\eqref{eq:mixing_para_u3} and Eq.~\eqref{eq:invariants_u3}. For the remaining two automorphisms $\mathfrak{u}_{10}$ and $\mathfrak{u}_{11}$, the model can only contain the unit representation $\mathbf{1}_{0, 0}$ and the triplet representation $\mathbf{3}_{\frac{n}{2}, \frac{n}{2}}$ with even $n$, if the corresponding CP transformation is imposed as a symmetry. The representation matrices of the generators $c$ and $d$ are given by
\begin{equation}
\rho_{\mathbf{3}_{\frac{n}{2}, \frac{n}{2}}}(c)=\text{diag}\left(-1, -1, 1\right),\qquad \rho_{\mathbf{3}_{\frac{n}{2}, \frac{n}{2}}}(d)=\text{diag}\left(1, -1, -1\right)\,,
\end{equation}
which yields $\rho_{\mathbf{3}_{\frac{n}{2}, \frac{n}{2}}}(c^2)=\rho_{\mathbf{3}_{\frac{n}{2}, \frac{n}{2}}}(d^2)=1$. Therefore $\mathbf{3}_{\frac{n}{2}, \frac{n}{2}}$ is not a faithful representation of $\Delta(3n^2)$ for $n>2$, and all the representations matrices form an $A_4$ group. The same lepton mixing patterns as those of $A_4$ flavor symmetry would be predicted.

\section{\label{sec:conclusion}Conclusion}

In the present paper, we have studied the possible lepton mixing patterns which can be achieved from the $\Delta(3n^2)$ family symmetry and generalized CP. Since the $\Delta(3n^2)$ groups don't have class-inverting automorphism except the two lowest order ones $\Delta(3\cdot1^2)\cong Z_3$ and $\Delta(3\cdot2^2)\cong A_4$, in generic settings it is not possible to define a proper generalized CP transformation compatible with the $\Delta(3n^2)$ family symmetry. However, if a specific model contains only a subset of irreducible representations which are mapped into their complex conjugates under the action of an automorphism $\mathfrak{u}$, then one can impose the generalized CP transformation corresponding to $\mathfrak{u}$ being a symmetry. By scrutinizing the consistency equations for the generators $a$ and $c$, in the case of $n\neq3\, \mathbb{Z}$, we find it is sufficient to consider only three types of automorphisms $\mathfrak{u}_1$, $\mathfrak{u}_2$ and $\mathfrak{u}_3$ with $\left(a, c\right)\stackrel{\mathfrak{u_1}}{\longmapsto}\left(a^{-1}, c^{-1}\right)$, $\left(a, c\right)\stackrel{\mathfrak{u_2}}{\longmapsto}\left(a^{-1}, c\right)$ and $\left(a, c\right)\stackrel{\mathfrak{u_3}}{\longmapsto}\left(a, c^{-1}\right)$. The required field content and the explicit form of the corresponding CP transformations are identified. In the case of $n=3\, \mathbb{Z}$, an additional eight generalized CP transformations could be introduced, as shown in Table~\ref{tab:GCP_n=3N}.

For each possible generalized CP transformation, we have studied the mixing patterns arising from the residual symmetry $G_{l}=Z_{3}$ in the charged lepton sector and $Z_2\times CP$ in the neutrino sector. We find that the PMNS matrix is always predicted to be the trimaximal pattern. As a consequence, the sum rule $3\sin^2\theta_{12}\cos^2\theta_{13}=1$ is fulfilled and the solar mixing angle has a lower limit given by $\sin^2\theta_{12}>1/3$ which can be tested in near future neutrino oscillation experiments. To be specific, if the generalized CP transformations relate to automorphisms mapping the generator $a$ into $n^2C_2$ or $\frac{n^2}{3}C^{(\tau)}_{2}$ with $\tau=0, 1, 2$, the PMNS matrix would be of the form of Eq.~\eqref{eq:PMNS_u1}. Three mixing angles are strongly correlated, both the Dirac phase $\delta_{CP}$ and the Majorana phase $\alpha_{31}$ are conserved while another Majorana phase $\alpha_{21}$ can take a set of discrete values. For the CP transformations defined by automorphisms mapping the generator $a$ into $n^2C_3$ or $\frac{n^2}{3}C^{(\tau)}_3$, the PMNS matrix is fixed to be of the form of Eq.~\eqref{eq:PMNS_u3}, where the expressions for $\varphi$ and $\rho$ vary with the residual symmetries. In this case, all the three CP phases depend on the free parameter $\theta$. They can be neither trivial nor maximal.

Finally we have discovered that the mixing patterns presented here for $\Delta(3n^2)$ are identical to those which have been obtained in the semidirect approach for the $\Delta(6n^2)$ family symmetry~\cite{Hagedorn:2014wha,Ding:2014ora}. The residual symmetries analyzed for the CP transformations corresponding to the automorphisms $\mathfrak{u}_2$ and $\mathfrak{u}_3$ are identified as the cases IV and case III of~\cite{Ding:2014ora} respectively. We emphasise that this identification is a new result that has not appeared in the literature until now. For the most interesting automorphism $\mathfrak{u}_3$, leading to non-trivial Dirac CP phase, corresponding to Eqs.~\eqref{eq:PMNS_u3}, \eqref{eq:mixing_para_u3}, we have presented the results for the first time as contour plots in the plane of $\theta$ and $\varphi$, as shown in Figures \ref{fig:contour_angles}, \ref{fig:contour_CP}.

\section*{Acknowledgements}

G.J.D is supported by the National Natural Science Foundation of China under Grant Nos. 11275188, 11179007 and 11522546. SK acknowledges support from the STFC Consolidated ST/J000396/1 grant and the European Union FP7 ITN-INVISIBLES (Marie Curie Actions, PITN- GA-2011- 289442).

\appendix

\renewcommand{\theequation}{\thesection\arabic{equation}}
\cleqn

\section{\label{sec:appendix_A}Clebsch-Gordan coefficients of $\Delta(3n^2)$ group}

As shown in section~\ref{sec:group_theory}, in the case of $n\neq3\, \mathbb{Z}$, the $\Delta(3n^2)$ group has three singlet representations $\mathbf{1}_{r}$ and $(n^2-1)/3$ triplet representations $\mathbf{3}_{(k,l)}$, where $r=0, 1, 2$ and $k, l=0, 1, \ldots, n-1$ with $(k, l)\neq(0,0)$. The Kronecker products of two irreducible representations are:
\begin{eqnarray}
\nonumber
\mathbf{1}_{r}\otimes\mathbf{1}_{r^{\prime}}&=&\mathbf{1}_{r+r^{\prime}\hskip-0.08in\pmod{3}},\\
\nonumber\mathbf{1}_{r}\otimes\mathbf{3}_{(k,l)}&=&\mathbf{3}_{(k, l)},\\
\nonumber\mathbf{3}_{(k, l)}\otimes\mathbf{3}_{(k^{\prime}, l^{\prime})}&=&\delta_{\mbox{\tiny$\left(\begin{array}{c}
    k^{\prime} \\
    l^{\prime}
    \end{array}\right)$},\widetilde{\mbox{\tiny$\left(\begin{array}{c}
    -k \\
    -l
    \end{array}\right)$}}}(\mathbf{1}_0\oplus\mathbf{1}_1\oplus\mathbf{1}_2)\\
    &&\quad\oplus\mathbf{3}_{\mbox{\tiny$\left(\begin{array}{c}
    k^{\prime}+k\\
    l^{\prime}+l
    \end{array}\right)$}}\oplus\mathbf{3}_{\mbox{\tiny$\left(\begin{array}{c}
    k^{\prime}-k-l\\
    l^{\prime}+k
    \end{array}\right)$}}\oplus\mathbf{3}_{\mbox{\tiny$\left(\begin{array}{c}
    k^{\prime}+l\\
    l^{\prime}-k-l
    \end{array}\right)$}}\,.
\end{eqnarray}
For $n=3\, \mathbb{Z}$, the $\Delta(3n^2)$ group has nine singlet representations $\mathbf{1}_{r, s}$ and $(n^2-3)/3$ triplet representations  $\mathbf{3}_{(k,l)}$, where $r, s=0, 1, 2$ and $k, l=0, 1, \ldots, n-1$ with $(k, l)\neq(0,0)$, $(n/3, n/3)$, $(2n/3, 2n/3)$. The Kronecker products are
\begin{eqnarray}
\nonumber\mathbf{1}_{r, s}\otimes\mathbf{1}_{r^{\prime}, s^{\prime}}&=&\mathbf{1}_{r+r^{\prime}\hskip-0.08in\pmod{3},~ s+s^{\prime}\hskip-0.08in\pmod{3}},\\
\nonumber\mathbf{1}_{r, s}\otimes\mathbf{3}_{(k,l)}&=&\mathbf{3}_{\widetilde{\mbox{\tiny $\left(\begin{array}{c}
    k+sn/3\\
    l+sn/3
    \end{array}
    \right)$}}},\\
\nonumber\mathbf{3}_{(k, l)}\otimes\mathbf{3}_{(k^{\prime}, l^{\prime})}&=&\sum_{s=0}^{2}\delta_{\mbox{\tiny$\left(\begin{array}{c}
    k^{\prime} \\
    l^{\prime}
    \end{array}\right)$},\widetilde{\mbox{\tiny$\left(\begin{array}{c}
    -k+sn/3 \\
    -l+sn/3
    \end{array}\right)$}}}(\mathbf{1}_{0,s}\oplus\mathbf{1}_{1,s}\oplus\mathbf{1}_{2,s})\\
    &&\quad\oplus\mathbf{3}_{\mbox{\tiny$\left(\begin{array}{c}
    k^{\prime}+k\\
    l^{\prime}+l
    \end{array}\right)$}}\oplus\mathbf{3}_{\mbox{\tiny$\left(\begin{array}{c}
    k^{\prime}-k-l\\
    l^{\prime}+k
    \end{array}\right)$}}\oplus\mathbf{3}_{\mbox{\tiny$\left(\begin{array}{c}
    k^{\prime}+l\\
    l^{\prime}-k-l
    \end{array}\right)$}}\,.
\end{eqnarray}
In the following, we shall present all the Clebsch-Gordan coefficients in the form of $\alpha\otimes\beta$ in our chosen basis, $\alpha_i$ denotes the element of the left base vectors $\alpha$, and $\beta_i$ is the element of the right base vectors $\beta$.

\begin{description}
\item[\textbf{(i)}] {$n\neq3\, \mathbb{Z}$}

\begin{itemize}[labelindent=-0.7em, leftmargin=0.1em]

\item{$\mathbf{1}_{r}\otimes\mathbf{1}_{r^{\prime}}=\mathbf{1}_{r+r^{\prime}\hskip-0.05in\pmod{3}}$}

\begin{equation}
\mathbf{1}_{r+r^{\prime}\hskip-0.08in\pmod{3}}=\alpha\beta\,.
\end{equation}

\item{$\mathbf{1}_{r}\otimes\mathbf{3}_{(k,l)}=\mathbf{3}_{(k, l)}$}

\begin{equation}
\mathbf{3}_{(k, l)}=\left(\begin{array}{c}
\alpha\beta_1\\
\alpha\beta_2\\
\alpha\beta_3\\
\end{array}\right)\,.
\end{equation}

\item{$\mathbf{3}_{(k, l)}\otimes\mathbf{3}_{(k^{\prime}, l^{\prime})}=\delta_{\mbox{\tiny$\left(\begin{array}{c}
    k^{\prime} \\
    l^{\prime}
    \end{array}\right)$},\widetilde{\mbox{\tiny$\left(\begin{array}{c}
    -k \\
    -l
    \end{array}\right)$}}}(\mathbf{1}_0\oplus\mathbf{1}_1\oplus\mathbf{1}_2)\oplus\mathbf{3}_{\mbox{\tiny$\left(\begin{array}{c}
    k^{\prime}+k\\
    l^{\prime}+l
    \end{array}\right)$}}\oplus\mathbf{3}_{\mbox{\tiny$\left(\begin{array}{c}
    k^{\prime}-k-l\\
    l^{\prime}+k
    \end{array}\right)$}}\oplus\mathbf{3}_{\mbox{\tiny$\left(\begin{array}{c}
    k^{\prime}+l\\
    l^{\prime}-k-l
    \end{array}\right)$}}$}

\begin{equation}
\hskip-0.25in\mathbf{3}_{\mbox{\footnotesize$\left(\begin{array}{c}
k^{\prime}+k\\
l^{\prime}+l
\end{array}\right)$}}=\left(\begin{array}{c}
\alpha_1\beta_1\\
\alpha_2\beta_2\\
\alpha_3\beta_3
\end{array}\right),~~ \mathbf{3}_{\mbox{\footnotesize$\left(\begin{array}{c}
k^{\prime}-k-l\\
l^{\prime}+k
\end{array}\right)$}}=\left(\begin{array}{c}
\alpha_2\beta_1\\
\alpha_3\beta_2\\
\alpha_1\beta_3
\end{array}\right),~~
\mathbf{3}_{\mbox{\footnotesize$\left(\begin{array}{c}
k^{\prime}+l\\
l^{\prime}-k-l
\end{array}\right)$}}=\left(\begin{array}{c}
\alpha_3\beta_1\\
\alpha_1\beta_2\\
\alpha_2\beta_3
\end{array}\right)\,.
\end{equation}

If $(k^{\prime}, l^{\prime})=\widetilde{\left(-k, -l\right)}$, then one of the above three triplet representations would be absent since the representation $\mathbf{3}_{0, 0}$ is reducible. For a generic triplet field $\varphi=\left(\varphi_1, \varphi_2, \varphi_3\right)^{T}$ transforming as $\mathbf{3}_{0,0}$, it can be reducible into three singlet $\mathbf{1}_{0}$, $\mathbf{1}_1$ and $\mathbf{1}_{2}$, i.e., we have $\mathbf{3}_{(0, 0)}=\mathbf{1}_{0}\oplus\mathbf{1}_{1}\oplus\mathbf{1}_{2}$ with
\begin{equation}
\mathbf{1}_{r}=\varphi_1+\omega^{-r}\varphi_2+\omega^{r}\varphi_3,~~\text{with}~~r=0, 1, 2\,.
\end{equation}
For example, in the case of $(k^{\prime}, l^{\prime})=(-k, -l)$, the first triplet $\mathbf{3}_{\mbox{\tiny$\left(\begin{array}{c}
    k^{\prime}+k\\
    l^{\prime}+l
    \end{array}\right)$}}$ is absent and it is replaced by three singlets $\alpha_1\beta_1+\alpha_2\beta_2+\alpha_3\beta_3$, $\alpha_1\beta_1+\omega^2\alpha_2\beta_2+\omega\alpha_3\beta_3$ and $\alpha_1\beta_1+\omega\alpha_2\beta_2+\omega^2\alpha_3\beta_3$ instead.
\end{itemize}

\item[\textbf{(ii)}] {$n=3\, \mathbb{Z}$}

\begin{itemize}[labelindent=-0.7em, leftmargin=0.1em]

\item{$\mathbf{1}_{r, s}\otimes\mathbf{1}_{r^{\prime}, s^{\prime}}=\mathbf{1}_{r+r^{\prime}\hskip-0.05in\pmod{3},~ s+s^{\prime}\hskip-0.05in\pmod{3}}$}

\begin{equation}
\mathbf{1}_{r+r^{\prime}\hskip-0.08in\pmod{3},~ s+s^{\prime}\hskip-0.08in\pmod{3}}=\alpha\beta\,.
\end{equation}

\item{$\mathbf{1}_{r, s}\otimes\mathbf{3}_{(k,l)}=\mathbf{3}_{\widetilde{\mbox{\tiny $\left(\begin{array}{c}
    k+sn/3\\
    l+sn/3
    \end{array}
    \right)$}}}$}

\begin{equation}
\mathbf{3}_{\widetilde{\mbox{\footnotesize $\left(\begin{array}{c}
    k+sn/3\\
    l+sn/3
    \end{array}
    \right)$}}}=\left(\begin{array}{c}
\alpha\beta_1\\
\alpha\beta_2\\
\alpha\beta_3\\
\end{array}\right)\,.
\end{equation}

\item{$\mathbf{3}_{(k, l)}\otimes\mathbf{3}_{(k^{\prime}, l^{\prime})}=\sum_{s=0}^{2}\delta_{\mbox{\tiny$\left(\begin{array}{c}
    k^{\prime} \\
    l^{\prime}
    \end{array}\right)$},\widetilde{\mbox{\tiny$\left(\begin{array}{c}
    -k+sn/3 \\
    -l+sn/3
    \end{array}\right)$}}}(\mathbf{1}_{0,s}\oplus\mathbf{1}_{1,s}\oplus\mathbf{1}_{2,s})\oplus\mathbf{3}_{\mbox{\tiny$\left(\begin{array}{c}
    k^{\prime}+k\\
    l^{\prime}+l
    \end{array}\right)$}}\oplus\mathbf{3}_{\mbox{\tiny$\left(\begin{array}{c}
    k^{\prime}-k-l\\
    l^{\prime}+k
    \end{array}\right)$}}\oplus\mathbf{3}_{\mbox{\tiny$\left(\begin{array}{c}
    k^{\prime}+l\\
    l^{\prime}-k-l
    \end{array}\right)$}}$}

\begin{equation}
\label{eq:3times3_N3}\hskip-0.25in\mathbf{3}_{\mbox{\footnotesize$\left(\begin{array}{c}
k^{\prime}+k\\
l^{\prime}+l
\end{array}\right)$}}=\left(\begin{array}{c}
\alpha_1\beta_1\\
\alpha_2\beta_2\\
\alpha_3\beta_3
\end{array}\right),~~ \mathbf{3}_{\mbox{\footnotesize$\left(\begin{array}{c}
k^{\prime}-k-l\\
l^{\prime}+k
\end{array}\right)$}}=\left(\begin{array}{c}
\alpha_2\beta_1\\
\alpha_3\beta_2\\
\alpha_1\beta_3
\end{array}\right),~~
\mathbf{3}_{\mbox{\footnotesize$\left(\begin{array}{c}
k^{\prime}+l\\
l^{\prime}-k-l
\end{array}\right)$}}=\left(\begin{array}{c}
\alpha_3\beta_1\\
\alpha_1\beta_2\\
\alpha_2\beta_3
\end{array}\right)\,.
\end{equation}

As the representation $\mathbf{3}_{sn/3, sn/3}$ with $s=0, 1, 2$ is reducible in this case, one or three of the triplet representations in Eq.~\eqref{eq:3times3_N3} would be absent if $\mbox{\tiny$\left(\begin{array}{c}
    k^{\prime} \\
    l^{\prime}
    \end{array}\right)$}=\widetilde{\mbox{\tiny$\left(\begin{array}{c}
    -k+sn/3 \\
    -l+sn/3
    \end{array}\right)$}}$ is fulfilled. For a generic triplet field $\varphi=\left(\varphi_1, \varphi_2, \varphi_3\right)^{T}$ transforming as $\mathbf{3}_{sn/3, sn/3}$, it can be reduced into three singlet $\mathbf{1}_{0, s}$, $\mathbf{1}_{1, s}$ and $\mathbf{1}_{2, s}$ with
\begin{equation}
\mathbf{3}_{sn/3, sn/3}=\mathbf{1}_{0, s}\oplus \mathbf{1}_{1, s}\oplus \mathbf{1}_{2, s}\,,
\end{equation}
where
\begin{equation}
\mathbf{1}_{r, s}=\varphi_1+\omega^{-r}\varphi_2+\omega^{r}\varphi_3,~~\text{with}~~r=0, 1, 2\,.
\end{equation}
For example, in the case of $(k^{\prime}, l^{\prime})=(-k, -l)=(0, \pm n/3)$, $(\mp n/3, 0)$ or $(\pm n/3, \mp n/3)$, one obtain nine one-dimensional representations from the product of two triplet as follows,
\begin{eqnarray}
\nonumber\mathbf{1}_{r,0}&=&\alpha_1\beta_1+\omega^{-r}\alpha_2\beta_2+\omega^{r}\alpha_3\beta_3,\\
\nonumber\mathbf{1}_{r,1} (\mathrm{or}~\mathbf{1}_{r, 2})&=&\alpha_2\beta_1+\omega^{-r}\alpha_3\beta_2+\omega^{r}\alpha_1\beta_3,\\ \mathbf{1}_{r,2} (\mathrm{or}~\mathbf{1}_{r, 1})&=&\alpha_3\beta_1+\omega^{-r}\alpha_1\beta_2+\omega^{r}\alpha_2\beta_3\,.
\end{eqnarray}

\end{itemize}

\end{description}

\end{document}